\documentclass[sigplan,10pt,nonacm]{acmart}
\usepackage{booktabs}   %
\usepackage{subcaption} %
\usepackage{tikz}
\usepackage{amsmath}
\usepackage{multicol}

\usepackage{filecontents}

\usepackage{todonotes}
\usepackage{comment}

\usepackage{multirow}

\usepackage[linesnumbered,ruled,vlined,commentsnumbered]{algorithm2e}
\SetArgSty{textnormal}
\SetAlFnt{\scriptsize}
\SetAlCapFnt{\scriptsize}
\SetAlCapNameFnt{\scriptsize}
\DontPrintSemicolon
\usepackage{listings}
\usepackage{subcaption}
\usepackage{paralist}
\usepackage{ulem}

\begin{document}

\title[Bundled References]{Bundled References: An Abstraction for Highly-Concurrent Linearizable Range Queries}         %
\author{Jacob Nelson}
\affiliation{
  \department{Computer Science and Engineering}              %
  \institution{Lehigh University}            %
  \country{USA}                    %
}
\email{jjn217@lehigh.edu}          %

\author{Ahmed Hassan}
\affiliation{
  \department{Computer Science and Engineering}              %
  \institution{Lehigh University}            %
  \country{USA}                    %
}
\email{ahh319@lehigh.edu}          %

\author{Roberto Palmieri}
\affiliation{
  \department{Computer Science and Engineering}              %
  \institution{Lehigh University}            %
  \country{USA}                    %
}
\email{palmieri@lehigh.edu}          %

\begin{abstract}
We present bundled references, a new building block to provide linearizable range query operations for highly concurrent linked data structures. Bundled references allow range queries to traverse a path through the data structure that is consistent with the target atomic snapshot and is made of the minimal amount of nodes that should be accessed to preserve linearizability. We implement our technique into a skip list, a binary search tree, and a linked list data structure. Our evaluation reveals that in mixed workloads, our design improves upon the state-of-the-art techniques by 3.9x for a skip list and 2.1x for a binary search tree. We also integrate our bundled data structure into the DBx1000 in-memory database, yielding up to 20\% gain over the same competitors.
\end{abstract}


\maketitle

\section{Introduction}

Iterating over a collection of elements to return those that
fall within a contiguous range (also known as a \textit{range query} operation) is without contest an essential feature for data repositories.
In addition to database management systems, which historically deploy support for range queries (through predicate reads or writes), recent key-value stores (e.g., RocksDB~\cite{facebook2020rocksdb} and others~\cite{DBLP:conf/sigcomm/EscrivaWS12,DBLP:conf/eurosys/MaoKM12,silk,pebblesdb,google2019leveldb}) enrich their traditional PUT and GET APIs to include range query operations.

With the high-core-count era in full swing, providing high-performance range query operations that execute concurrently with modifications is challenging.
On the one hand, ensuring strong correctness guarantees of range queries, such as linearizablity~\cite{linearizability}, requires that they observe a consistent snapshot of the collection regardless of any concurrent update that may take place. On the other hand, since range queries are naturally read-only operations, burdening them with synchronization steps to achieve strong correctness guarantees may significantly deteriorate their performance.

To address this trade-off, existing solutions in literature either assume relaxed guarantees for range queries~\cite{java-util-concurrent-lib,rlu} or sacrifice providing high-performance
under highly concurrent workloads, namely when hundreds of threads concurrently perform a mix of updates on single elements (so called primitive operations) and scans over a range of elements~\cite{ebr-rq,rlu,snapcollector}.

In this paper we propose \textit{bundled references}, a new building block to design linearizable concurrent linked data structures (e.g., skip lists) optimized to scale up performance of range query operations executing concurrently with update operations. %
The core innovation behind bundled reference lies in adapting the design principle of Multi Version Concurrency Controls (MVCC)~\cite{wu2017empirical} to generic linked data structures, and improving it by eliminating the overhead of determining the appropriate version to be returned.
Bundled references achieve that by augmenting each link in a data structure with a record of its previous values, each of which is tagged with a timestamp reflecting the point in (logical) time when the operation that generated that link occurred. In other words, we associate timestamps to references connecting data structure elements instead of to the pointed elements.

The bundled reference building block enables the following properties of the data structure:
\begin{compactitem}
\item Range query operations are linearized when they start, which helps reduce the interference with ongoing and subsequent update operations;
\item Each thread performing a range query only traverses the \textit{minimal} amount of nodes in the range, 
regardless of concurrent updates. The minimality property precludes a thread from scanning multiple versions to find the one consistent with its linearizable snapshot.
\item Data structure traversals, including those of contains and update operations, do not require any special treatment, permitting optimizations such as wait-free~\cite{wait-freedom} traversals in the lazy data structure patterns;
\item State-of-art reclamation techniques (e.g., EBR~\cite{ebr}) can be easily integrated into the bundled references to reclaim data structure elements, which minimizes the space overhead of bundling, making it practical.
\end{compactitem}

By leveraging the bundled references, we develop three relevant ordered implementations of a Set, namely a linked list, a skip list, and a binary search tree (BST).
While the linked list is a convenient data structure to illustrate all the details of our design that favors range query operations, the skip list and the BST are high-performance data structures widely used in systems (such as database indexes) where predicate reads
are predominant.

In these new data structure implementations we replace the existing links with bundled references to provide linearizable range queries.
The history of a link between nodes (called a \textit{bundle}) is consistently updated every time a successful modification to the data structure occurs.
A range query uses the bundles in each node to perform its scan on the data structure, following a path made of the latest links marked with a timestamp lower than (or equal to) the operation's starting timestamp.

We evaluate our bundled data structures against three alternative techniques to provide linearizable range queries, namely read-log-update (RLU) and two variants of a solution based on epoch-based reclamation (EBR-RQ and EBR-RQ-LF). In a mixed workload, bundling allows for up to 3.9x and 2.1x improvement over the closest comptetitors.
We also find that we outperform the competitors at high thread counts (i.e., 192 threads), in nearly all cases. Further, bundling achieves a more consistent performance profile across different configurations than our competitors, whose design choices lead them to prefer specific workloads.
Finally, we integrate both the skip list and BST as indexes in the DBx1000 in-memory database and test them using the TPC-C benchmark, finding that bundling provides 20\% better performance than the next best competitor at high numbers of threads.

To the best of our knowledge, bundling is the first approach that allows for range query operations over a linked data structure that traverse the minimal amount of nodes in their ranges without blocking concurrent update operations in the range.
The source code for our bundling technique and bundled data structures can be found at \url{https://zenodo.org/record/4402298}. Refer to the README, therewithin for info
\section{Related Work}
\label{sec:rel-work}

\textbf{Linearizable range queries.}
Existing work has focused on providing range queries through highly-specific data structure implementations~\cite{karytree,ctrie,catree,leaplist,DBLP:conf/ppopp/BronsonCCO10,kiwi}.
While recognizing their effectiveness,
their tight dependency on the data structure characteristics makes them difficult to extend to other structures, even if manually. 
The literature is also rich with highly effective concurrent data structure designs that lack range query support and cannot leverage the aforementioned data structure specific solutions to perform range queries. 
This motivates generalized solutions, which achieve linearizable range queries by applying the same technique to a variety of data structures~\cite{ebr-rq, rlu, snapcollector}.

Read-log-update (RLU)~\cite{rlu} is a technique in which writing threads keep a local log of updated objects, along with the logical timestamp when the update takes effect. 
When no reader requires the original version, the log is committed.
It extends read-copy-update (RCU)~\cite{rcu} to supports multiple object updates. 
Similar to Bundling, RLU's range queries are linearized at the beginning of their execution, after reading a global timestamp and fixing their view of the data structure.
However, in RLU, updates block while there are ongoing RLU protected operations,
as it only commits its changes after guaranteeing no operation will access the old version. 
Bundling minimizes write overhead because new entries are added while deferring the removal of outdated ones.

Snapcollector~\cite{snapcollector} also logs changes made to the data structure during an operations lifetime so that concurrent updates are observed.
A range query first announces its intention to snapshot the data structure by posting a reference to an object responsible for collecting updates.
It traverses as it would in a sequential setting, then checks a report of concurrent changes it may have missed.
The primary difference with respect to RLU is that range queries are linearized at the end of the operation, after disabling further reports.

Although the construction of Snapcollector is wait-free, this method may lead a range query to observe reports of changes that were already witnessed during its traversal.
Creating and announcing reports penalizes operation performance; not to mention the memory overhead required to maintain these reports.
Collectively, the cost of these characteristics is insurmountable and we experimentally verify that it is easily outperformed.
With our bundling approach, a range query visits nodes only once to produce its view of the data structure and is linearized at the operation's start.

An extension of Snapcollector enables snapshotting only a range of the data structure instead of all elements~\cite{chatterjee2017lock}. However, this approach continues to suffer many of the same pitfalls as the original design. In addition to these, concurrent range queries with overlapping key ranges are disallowed.

Arbel-Raviv and Brown~\cite{ebr-rq} build upon epoch-based memory reclamation (EBR) to provide linearizable range queries.
In this method, range query traversals leverage a global timestamp to determine if nodes, annotated with a timestamp by update operations, belong in their snapshot.
In order to preserve linearizability, remove operations announce their intention to delete a node before physically removing them and adding them to the list of to-be-deleted nodes, or limbo list. 
Range queries scan the data structure, the announced deletions, and limbo list to determine which nodes to include in their view; potentially resulting in a situation where nodes are observed multiple times.
Since range queries' atomic updates to the timestamp conflict, the design also prioritizes update-mostly workloads.

Our bundling approach enhances performance of range queries by allowing them to traverse the minimal number of nodes in the range without needing to validate its snapshot and eliminating contention on a shared global counter.

\textbf{MVCC.}
Multi-version concurrency control (MVCC) relies on timestamps to 
coordinate concurrent accesses to objects.
It is widely used in database management systems.
Many different implementations exist~\cite{ben2019multiversion, bernstein1983multiversion, neumann2015fast, larson2011high, lim2017cicada}; all rely on a multiversioned data repository where each shared object stores a list of versions,
and each version is tagged with a creation timestamp.
Transactions then read the versions of objects that are consistent with their execution.

The de facto standard for version storage in MVCC systems is to maintain a linked list of versions for each object that is probed during a read~\cite{wu2017empirical}, with
recent innovations targeting this particular aspect.
One example, Cicada~\cite{lim2017cicada}, implements optimistic MVCC by making a transaction install a \texttt{PENDING} version for every written object as the first step in its validation phase. 
Readers block until this version is no longer pending.
When the transaction either commits or aborts each \texttt{PENDING} version's status changes accordingly.

Similar to Cicada, bundling requires updates to install pending entries to notify other operations of an ongoing change; in contrast to Cicada's approach, updates are always successful.
Futhermore, pending entries only exist for a short duration surrounding the linearization point.

Another example, X-Engine~\cite{huang2019x}, is a highly optimized LSM-tree storage engine that
uses a skip list variant to optimize access to recent updates stored in memory.
A key is represented by a single node pointing to a list of versions.
Comparable approaches (e.g. LevelDB~\cite{google2019leveldb} and RocksDB~\cite{facebook2020rocksdb}), treat versions as independent nodes in the data structure. X-engine's solution optimizes for index layer traversals since the path no longer includes multiple versions of the same key. Nevertheless, the version list must still be scanned for the value consistent with the operation.
Unlike X-engine, bundled references ensures that the MVCC traverses \textit{only} those objects that belong to the transaction's atomic snapshot. 

\textbf{Persistent data structures.}
The Bundled reference abstraction is similar in spirit to the concept of \textit{fat nodes} in \textit{persistent data structures}~\cite{driscoll1986making}. 
In principle, persistent data structures are those which maintain all previous versions of the data structure. 
The ephemeral structure is the current state and the persistent structure encodes past ephermeral structures. 
The core difference is that bundling aims at providing efficient linearizable range queries in highly concurrent workloads, while persistent data structures are commonly used in functional programming languages to maintain theoretical requirements regarding object immutablility~\cite{okasaki1999purely, hickey2008clojure} and algorithms requiring reference to previous state~\cite{sarnak1986planar, chien2001efficient}.

%
%
%

%
%
%
%

%
%
%

%
%
%

%
%
%
%

%
%
%
%

%
%
%
%

%
\section{The Bundle Building Block}
\label{sec:overview}

The principal idea behind \textit{bundling} is the maintenance of a historical record of physical changes to the data structure so that range queries can traverse a consistent snapshot.
As shown in details below, the idiosyncrasy of bundling is that this historical record stores \textit{links} between data structure elements that are used by range query operations to rebuild the exact composition of a range at a given (logical) time.

Before detailing bundling, it is important to note that update operations are totally ordered using a global timestamp, named \texttt{globalTs},
which is incremented every time a modification to the data structure takes place (i.e., when an update operation reaches its linearization point).

Every link in the data structure is backed by a \textit{bundle}, 
implemented as a linked list of \textit{bundle entries} (Listing~\ref{lst:classes1}). Each bundle entry logs the value of the link and the value of \texttt{globalTs} at the time the link was added to the bundle.
Whenever an update operation reaches its linearization point, meaning when it is guaranteed to complete, it prepends a bundle entry consisting of the newest value of the link and the value of the global timestamp.
Because of that, the head of the bundle always reflects the link's latest value.

Since each link's history is preserved through the bundles, range queries simply need to read the global timestamp at the operation's outset and traverse the linked data structure using the newest values no larger than the observed global timestamp.
This design is inherently advantageous when pruning bundle entries. 
In fact, a bundle entry may be removed (or recycled)
if an entry is no longer the newest in the bundle and no range query needs it.
\begin{figure}[h]
    \centering
    \includegraphics[width=0.4\textwidth]{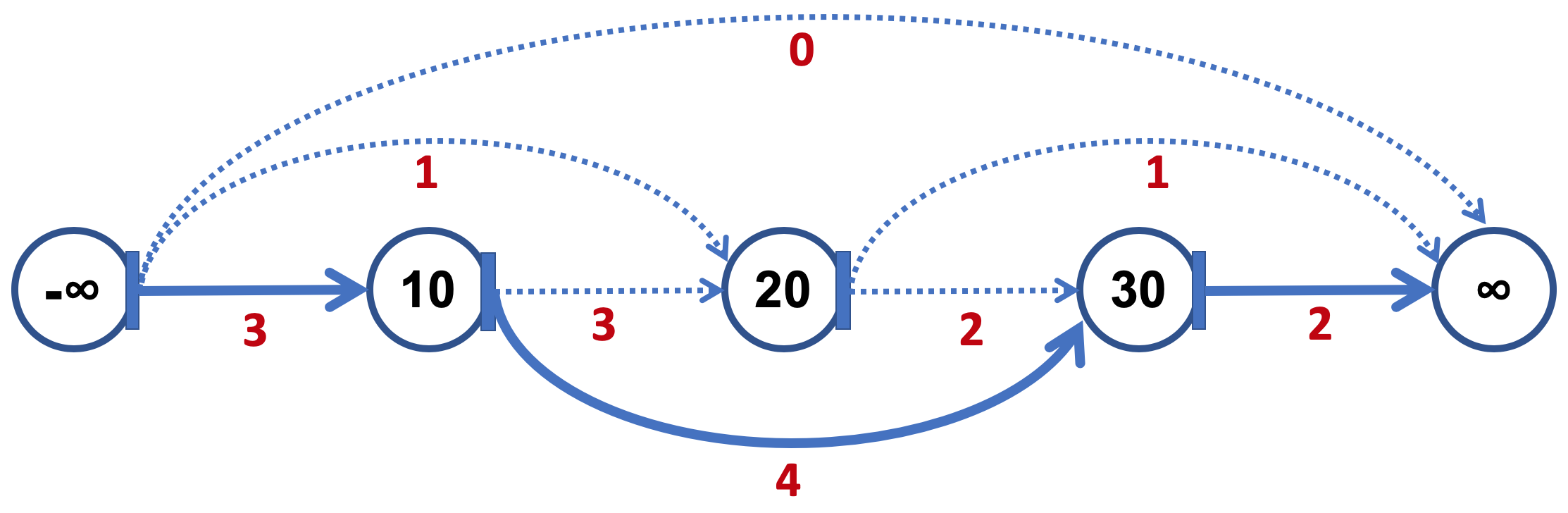}
    \caption{An example of using bundled references in a linked list. The path made of solid lines represents the state of the linked list after all update operations take place. Edges are labeled with their respective timestamps.}
    \label{fig:bundle-example}
\end{figure}

Figure~\ref{fig:bundle-example} shows an example on how bundles are deployed in a linked list.
As shown in the figure, the next pointer of each node is replaced by a bundle object that encapsulates the history of this next pointer. The figure shows the state of the linked list and its bundles after the following sequence of operations (assuming an initially empty linked list): \texttt{insert(20), insert(30), insert(10), remove(20)}.

To understand how this state is generated, we assume that the list is initialized with a single bundle reference whose timestamp is ``0'' (the initial value of \texttt{globalTs}), which connects its head and tail sentinel nodes. Inserting \texttt{20} does not replace this reference. Instead, it creates a new entry in the head's bundle with timestamp ``1'' pointing to the newly inserted node as well as an entry with the same timestamp in this new node pointing to the tail node. Similarly, inserting \texttt{30} and \texttt{10} adds new bundle entries with timestamps  ``2'' and ``3'', respectively. The last operation that removes \texttt{20} also does not replace any reference. Instead, it creates a new bundle entry in \texttt{10}'s bundle (with timestamp ``4'') that points to \texttt{30}, which reflects physically deleting \texttt{20} by making its predecessor node point to its next node.

Now assume that different range queries start at different times concurrently with those update operations. For clarity, we name a range query $R_i$ if it starts when \texttt{globalTs} is $i$, and for simplicity we assume its range matches the key range. Regardless of the times at which the different nodes are traversed, each range query is always able to traverse the proper snapshot of the list that reflects the time it started. For example, $R_0$ will skip any links in the range added after it started because all of them have timestamp greater than ``0''. Also, $R_3$ will observe \texttt{20} even if it reaches \texttt{10} after \texttt{20} is deleted. This is because in that case it will use the bundle entry whose timestamp is ``3'', which points to \texttt{20}.

The solid lines in the figure represent the most recent state of the linked list. Different insights can be inferred from this solid path. First, the references in this path are those with the largest timestamp in each bundle. This guarantees that any operation (including range queries) that starts after this steady state observes the most recent abstract state of the list. Second, once the reference with timestamp ``4'' is created, \texttt{20} becomes no longer reachable by any operation that will start later, because this operation will observe a timestamp greater than (or equal to) ``4''. Thus,
unreachable elements can be concurrently reclaimed.
In the following, to simplify
the description we assume that a bundle may hold an infinite number of entries, and no memory is freed.
Later in Section~\ref{sec:memreclamation}, and more in detail in the supplementary material, we address memory reclamation.

\subsection{Bundle Structure}
\label{sec:bundledref}

Generally, in order to deploy bundling each link in a data structure should use our bundled reference.
As an illustrative example, Listing~\ref{lst:classes2} 
shows how the \texttt{nextPtr} pointer in a linked list node is replaced with a bundled reference, which consists of the original reference (\texttt{newestNextPtr}) along with a bundle to record its history (\texttt{nextPtrBundle}).

\noindent\begin{minipage}{.21\textwidth}
\begin{lstlisting}[language=C++, frame=single, escapechar=|, basicstyle=\scriptsize,  numbers=left, stepnumber=1, numbersep=5pt, xleftmargin=10pt, framexleftmargin=8pt, label=lst:classes1, caption= Bundle.]
timestamp_t globalTs;
class BundleEntry {|\label{line:bundleentry1}|
    Node * ptr;
    timestamp_t ts;
    BundleEntry * next;
}|\label{line:bundleentry2}|
class Bundle {|\label{line:bundle1}|
    BundleEntry * head;
}|\label{line:bundle2}|

\end{lstlisting}
\end{minipage}\hfill
\begin{minipage}{.24\textwidth}
\begin{lstlisting}[language=C++, frame=single, escapechar=|, basicstyle=\scriptsize,  numbers=left, stepnumber=1, numbersep=5pt, xleftmargin=5pt, framexleftmargin=10pt, label=lst:classes2, caption= Linked List Node.]
class Node {|\label{line:node1}|
    key_t key;
    val_t val;
    lock_t lock;
    bool deleted;
    // The bundled reference.
    Node * newestNextPtr;
    Bundle nextPtrBundle;
}|\label{line:node2}|

\end{lstlisting}
\end{minipage}

%
%
Our bundle is a collection of entries sorted by timestamp to facilitate determining which reference to use during range queries.
Each bundle entry contains the updated pointer value, \texttt{ptr}, the timestamp associated with this value, \texttt{ts}, and a pointer to the next bundle entry, \texttt{next}.
As we detail in the next section, since update operations annotate bundle entries using a monotonically increasing timestamp (\texttt{globalTs}), new bundle entries will always have a timestamp larger than all other entries in the bundle.
Hence, bundle entries are sorted by \texttt{ts}.

A bundled reference consistently replicates the newest entry of the bundle (i.e., \texttt{newestNextPtr}). This is done because in our bundled data structures, primitive operations (i.e., add, remove, and contains) should not incur overhead due to the existence of the bundles.
As a positive side effect of this decision, all traversals done to reach the desired elements of the data structure, including those performed by range queries to enter the range, execute quickly without accessing the bundled references, as in a non-bundled data structure.
Meanwhile, range queries rely solely on bundles upon entering their range.

\subsection{Bundles and Update Operations}
\label{sec:updates}

Generally speaking, an update operation has two phases.
The operation first traverses the data structure to reach the desired location where the operation should take place, then performs the necessary changes.
Bundling involves augmenting only the act of changing pointers, not the traversal.
\begin{algorithm}
\KwIn{linAddr, linNewVal, bundles, ptrs}
\Begin{
\For{($b$,$p$) \textbf{in} ($bundles$, $ptrs$)}{
    PrepareBundle($b$, $p$) \\
    \label{line:prepare}
}
$ts \leftarrow $ AtomicFetchAndAdd(\&$globalTs$, 1) + 1 \\
\label{line:fetchaddts}
$*linAddr \leftarrow linNewVal$ \tcc*[r]{Linearization point.}
\label{line:linearization}
\For{$b$ in $bundles$} {
    FinalizeBundle($b$, $ts + 1$) \\
    \label{line:finalize}
}
}
\caption{LinearizeUpdateOperation}
\label{algo:linupdateop}
\end{algorithm}
\begin{algorithm}
\KwIn{bundle, ptr}
\Begin{
    $newEntry \leftarrow$ \textbf{new} $BundleEntry$ \\
    $newEntry.ptr \leftarrow ptr$ \\
    $newEntry.ts \leftarrow$ PENDING\_TS \\
    \label{line:pending}
    \While{true}{
        $expected \leftarrow bundle.head$ \\
        $newEntry.next \leftarrow expected$ \\
        \lWhile{$currEntry.ts = $ PENDING\_TS}{\textbf{end}}
        \label{line:updatewaitpendingbundle}
        \If{AtomicCompareAndSwap(\&$bundle.head$, $expected$, $newEntry$)}{
        \label{line:preparecas}
            \Return \\
        }
    }
}
\caption{PrepareBundle}
\label{algo:prepare}
\end{algorithm}

The role of bundling with respect to updates is to reflect the changes observable at the operation's completion so that range queries may see a consistent view of the data structure. 
This is performed through four crucial steps (Algorithm~\ref{algo:linupdateop}).
First, bundles 
are prepared by atomically prepending a new bundle entry in a \textit{pending} state (Line~\ref{line:prepare}).
After preparing the bundles, \texttt{globalTs} is atomically fetched and incremented and its new value is stored locally (Line~\ref{line:fetchaddts}). 
Next, the linearization point is executed (Line~\ref{line:linearization}), making the update visible to other primitive operations.
Lastly, pending bundle entries are finalized by annotating them with the newly incremented timestamp (Line~\ref{line:finalize}).

Initializing new bundle entries in a pending state (Line~\ref{line:pending} of Algorithm~\ref{algo:prepare}) is needed because range queries must wait until pending bundle entries are finalized to guarantee that they do not miss a concurrent update that should be included in their snapshot (see the example at the end of Section~\ref{sec:rangequeries}).
Additionally, concurrent updates attempting to modify a currently pending bundle are blocked until the ongoing update is finalized (Line~\ref{line:updatewaitpendingbundle} of 
Algorithm~\ref{algo:prepare}). This is done so that concurrent updates to the same node are properly ordered by timestamp.
It is also possible to address this problem by assuming that all nodes whose bundles will change are locked. 
We choose not to do so to make our design independent of data structure specific optimizations (see Section~\ref{sec:lazylist}).

\subsection{Bundles and Range Query Operations}
\label{sec:rangequeries}

Much like updates, a range query consists of two phases.
First, it traverses the data structure to reach the entry point to its range.
Next, it scans the range node by node to collect its snapshot.
These two phases are represented in Algorithm~\ref{algo:rangequery} by the functions \texttt{GetFirstNodeInRange}
and \texttt{GetNext},
respectively.
These two functions are data structure specific and their details are discussed in the subsequent sections.

Before explaining how range query operations perform in a bundled data structure, let us define a bundle entry to \textit{satisfy} a timestamp $t$ if it was the newest entry in the bundle when the global timestamp equaled $t$.

\begin{algorithm}
\KwIn{low, high}
\KwOut{resultTuples}
\Begin{
    \While{$true$}{
        $resultTuples \leftarrow \emptyset$ \\
        $ts \leftarrow globalTs$ \\
        \label{line:readglobalts}
        ($curr$, $valid$) $\leftarrow$ GetFirstNodeInRange($low$, $high$, $ts$) \\
        \label{line:getfirstnodeinrange}
        \uIf{$!valid$}{ 
        \label{line:noliveentry}
            \textbf{continue} \tcc*[r]{No bundle entry found.}
            \label{line:continue}
        }
        \uElseIf{$curr != nullptr$}{
            $resultsTuples \leftarrow resultsTuples$ $\bigcup$ $(curr.key, curr.value)$ \\ 
            \While{$curr \leftarrow $ GetNext($curr$, $low$, $high$, $ts$)}{
                \label{line:getnext}
                $resultsTuples \leftarrow resultsTuples$ $\bigcup$ $(curr.key, curr.value)$ \\
            }
            \Return $resultTuples$ \tcc*[r]{Return snapshot.} 
            \label{line:rqreturn}
        }
        \Else{
            \Return $resultTuples$ \tcc*[r]{Range empty.}
            \label{line:emptyrange}
        }
    }
}
\caption{RangeQuery}
\label{algo:rangequery}
\end{algorithm}

A range query collects a consistent snapshot of the data structure by following only references created by operations linearized before
the outset of the range query.
This is accomplished by first reading the current value of \texttt{globalTs} into a local variable \texttt{ts} to fix its snapshot (Line~\ref{line:readglobalts}), then traversing to the start of the range using \texttt{GetFirstNodeInRange} (Line~\ref{line:getfirstnodeinrange}), and finally scanning the data structure based on \texttt{ts} using \texttt{GetNext} (Line~\ref{line:getnext}).

The \texttt{GetFirstNodeInRange} function consists of two key steps.
First, it performs an optimistic traversal of the data structure, without checking the bundles, until it reaches the node preceding the first node in the range. Then, it traverses using the bundles to return the first node in its range.

Note that \texttt{GetFirstNodeInRange}'s initial traversal without bundles reflects the most recent state of the data structure, and not necessarily the
snapshot that will be observed by the range query.
Thus, two cases should be considered here.
First, if the node preceding the range has been inserted after the range query started, then
no bundle entry satisfying \texttt{ts} exists.
Since the visibility of a consistent snapshot cannot be guaranteed if no entries satisfy \texttt{ts}, the traversal is invalid and the range query starts over (Algorithm~\ref{algo:rangequery}, Line~\ref{line:continue}) with the new value of the global timestamp.
The other possibility is that a bundle entry satisfying \texttt{ts} is found, in which case it is safe to start traversing using bundles.

The second phase of \texttt{GetFirstNodeInRange} traverses further only using bundles to enter the range by relying on the \texttt{DereferenceBundle} function.
Note that, it is possible for this traversal to visit nodes not in the range, typically removed after the range query started, before reaching the first node in the range.

When \texttt{GetFirstNodeInRange} successfully returns the first node in the range, it is appended to the results and then the next nodes are obtained by repeated calls to
\texttt{GetNext} (Line~\ref{line:getnext}).
This function must return the next node in the range, strictly accessing it through bundles, to ensure only nodes that can be included in the snapshot are traversed.
Internally, all implementations of \texttt{GetNext} will also use \texttt{DereferenceBundle}, which is described next.

Given a bundle and timestamp, the \texttt{DereferenceBundle} function
works as follows. A range query
first waits for a pending bundle entry to be finalized, if any;
then it scans the bundle for the first entry whose timestamp is less than or equal to \texttt{ts}, indicating whether one was found.

Blocking until the first entry is no longer pending is a necessary step to ensure that the range query waits for a concurrent update that is already linearized, but whose bundles are not yet finalized.
To illustrate this scenario, consider the concurrent execution of two threads: $T_1$, which inserts the element $x$, and $T_2$, which performs a contains operation on $x$ and then executes a range query whose range includes $x$.
Thread $T_1$ starts at timestamp $t$ and proceeds in isolation until it executes its linearization step,
at which point it stalls indefinitely before the bundles are finalized.
Then, $T_2$ executes its contains on $x$, returning \texttt{True} since the original linearization point has already been reached.
Without waiting for pending entries to be finalized, the subsequent range query (\texttt{ts}$=t+1$) would not return $x$ as belonging to the range, leading to an inconsistent view of the data structure.
Instead, a bundle entry's pending state stalls range queries until the ongoing update completes, allowing $T_2$'s range query to observe $x$ in this example.

A critical invariant that guarantees the correctness of bundling is that an active range query always finds its required path.
This is obvious if both deleted nodes and outdated bundle entries are never reclaimed, as the entire history of each reference is recorded.
In Section~\ref{sec:memreclamation} and in the supplementary material, we show how to preserve this invariant even when reclamation takes place.

\subsection{Bundles and Contains Operations}
\label{sec:contains}

Since bundles are only kept to ensure correct range queries, contains operations execute independently from the bundled references.
Consequently, implementations with optimized contains operations (e.g., lazy data structures with wait-free contains~\cite{lazylist, lazyskiplist, otblist, citrus}) can leverage bundling without restricting their execution to a more conservative progress guarantee, as we will see in Sections~\ref{sec:lazylist}-\ref{sec:citrus}.

\section{Bundled Linked List}
\label{sec:lazylist}

We now describe how to apply bundling to the well-known highly-concurrent lazy sorted linked list~\cite{lazylist}, 
which provides high performance through wait-free traversals and fine-grained locking updates.
We recall that Listing~\ref{lst:classes2} 
provides a full definition of member variables of its nodes.

The wait-free contains operation is the same as the original lazy linked list without bundling~\cite{lazylist}.
Therefore, a traversal 
uses \texttt{newestNextPtr}
to walk the list until it reaches the target key.
At this point, it returns a reference to the first node with key greater than or equal to the target and its predecessor.
It validates the current node returned from the traversal phase by checking its equivalence with the target key and if it is logically deleted. If validation passes, the contains operation returns \texttt{True}, otherwise it returns \texttt{False}. 

%
\begin{algorithm}
\KwIn{key, val}
\Begin{
    \While{true}{
       $pred, curr \leftarrow$ Traverse($key$) \\
       Lock($pred$) \\
       \label{line:lazylistinsertlock}
       \If{ValidateLinks($pred$, $curr$)}{
            \If{$curr.key == key$}{
                \Return false \\
            }
            $newNode \leftarrow$ \textbf{new} Node($key$, $val$) \\
            $newNode.newestNextPtr \leftarrow curr$ \\
            $bundles \leftarrow$ ($newNode.bundle$, $pred.bundle$) \\
            $ptrs \leftarrow$ ($curr$, $newNode$) \\
            LinearizeUpdateOperation(\&$pred.newestNextPtr$, $newNode$, $bundles$, $ptrs$) \\
            \label{line:lazylistinsertlinearize}
            Unlock($pred$) \\
            \Return true
       }
       Unlock($pred$) \\
   }
}
\caption{Insert operation of Bundled Linked List}
\label{algo:lazylistinsert}
\end{algorithm}

Similarly, insert operations (Algorithm~\ref{algo:lazylistinsert}) make use of the traversal to determine where the new node will be added.
After locking the predecessor, both current and predecessor nodes are validated by checking that they are not deleted and that no node was inserted between them.
If validation succeeds and the key does not already exist, a new node with its next pointer set to the appropriate node is created. If it fails, the nodes are unlocked and the operation restarts.

Up to this point we have followed the same procedure that a data structure without bundling would use.
The next step is to call \texttt{LinearizeUpdateOperation} to perform the four steps described in Section~\ref{sec:updates} to linearize an update operation in a bundled data structure: installing pending bundle entries, incrementing the global timestamp, performing the linearization point, and finalizing the bundles.
For an insertion, the bundles of the newly added node and its predecessor must be modified to reflect their new values and the timestamp of the operation.
The linearization point remains the moment that the predecessor's \texttt{newestNextPtr} is set to the new node.
Finally, the locks are released and the insertion returns \texttt{True}.

Note that in Algorithm~\ref{algo:lazylistinsert} we employ an optimization where only the predecessor is locked by insert operations (Line~\ref{line:lazylistinsertlock}). In~\cite{lazylist}, it has been proven that this optimization preserves linearizability.
However, this optimization reveals a subtle but important corner case that motivates the need for waiting for pending bundles to be finalized
(Line~\ref{line:updatewaitpendingbundle} of Algorithm~\ref{algo:prepare}).
Because the current node is not locked, it is possible that a concurrent update operation successfully locks the new node after it is reachable and before its bundles are finalized by the inserting operation.
This nefarious case is protected by first waiting for the ongoing insertion to finish to ensure the bundle remains ordered (see Section~\ref{sec:updates}).

Remove operations
follow a similar pattern by first traversing to the appropriate location, locking the nodes of interest, validating them (restarting if validation fails), removing the current node if its key matches the target key, then finally unlocking the nodes and returning.
Here, the removal is linearized when the node is logically deleted, not when the reference changes.
However, we reflect the predecessor's updated reference in the bundle since the physical removal of this node resides within the same critical section as its linearization point.
The removed node's bundle does not change because its \texttt{ptr} value reflects the physical state of the data structure immediately before the removal takes place.

%
%
%
%

%
%
%
%
%
%
%
%
%
%
%
%
%
%
%
%
%
%
%
%
%
%
%
%
%
%

We now turn our attention to the two functions required by range queries: \texttt{GetFirstNodeInRange} and \texttt{GetNext}.
Recall from Section~\ref{sec:rangequeries} that a range query must first traverse the list (without bundles) to the node pointing to the first node in the range, then enter the range by traversing only bundles.
For a linked list, we simply scan from the head until this node is found, then traverse using the bundles up to the first node in the range.
Traversing using \texttt{GetNext} is trivial since we simply return the node that satisfies \texttt{ts} in the bundle of the current node.
These two functions are used by Algorithm~\ref{algo:rangequery} to perform linearizable range query operations.

\textbf{Correctness}. Proving the linearizability of our bundled linked list is straightforward. The linearization point of (successful) update operations is the same as the original lazy linked list: insertion is linearized when the \texttt{newestNextPtr} of the predecessor node is changed, and removal is linearized when the node is logically deleted. Contains operations' linearization point also matches that of the original lazy linked list,
as they use \texttt{newestNextPtr} in their traversal and check for logical deletion when they reach the node.

A range query $R$ is linearized when it snapshots the global timestamp before starting its data structure traversal.
An update operation $U$ that is linearized before this point will always be observed by $R$ because only the following two situations can occur. First, $U$ completely finished before $R$ starts,
which means that the timestamps of the links added by $U$ are less than, or equal to, the snapshot taken by $R$.
Second, $U$ is concurrent with $R$ but executes its linearization step before $R$ starts. In this case, since the linearization step of $U$ is executed only after incrementing the global timestamp and changing the corresponding bundles into a pending state, $R$ will be blocked (if needed) until the pending states of such bundles are released and the links that reflect the updates made by $U$ are added with the proper timestamp. 

\textbf{Minimality of traversed nodes within range}. One of the powerful properties of the bundled linked list is that range queries traverse the minimum number of nodes in the range: starting from the first node in the range, the range query only scans the nodes that belong to its range. 
It is worth noting that this minimality would also hold for the traversal phase (before reaching the entry node to the range) if we would have used bundles from the beginning of the list. 
However, as we mentioned before, for performance reasons
we decide to avoid using bundles to reach the first node of the range, and instead we traverse through \texttt{newestNextPtr}.

\textbf{Space overhead}. Although space overhead may seem a concern, it is instead acceptable in practical deployments. Even without reclamation, insertions have an amortized constant overhead since each insert operation adds two bundle entries for each new node instead of adding one new link in the non-bundled lazy list. This means that a list of $n$ nodes (assuming no removals) will have a total of $2n$ bundle entries. Enabling reclamation reduces this number significantly.
If a cleanup operation is performed while all other threads are in a quiescent state, it is guaranteed to leave only one in each bundle (see Section~\ref{sec:memreclamation} and the supplementary material).
\section{Bundled Skiplist}
\label{sec:skiplist}

The second data structure where we apply bundling is the lazy skip list~\cite{lazyskiplist}, whose update operations use fine-grained locks and contains operations are wait-free, similar to the bundled linked list. 
In the following, we highlight the differences between the two designs. 

The first difference is that skip list consists of a bottom \textit{data layer} where data resides and a set of \textit{index layers} to accelerate traversal. 
Hence, given a target key, traversal returns a set of (\texttt{pred}, \texttt{curr}) pairs for both index and data layers, rather than a single pair.
If the target key exists in the data structure, then it also returns the highest level (\texttt{levelFound}) at which the node was found. 
A naive approach to bundling this skip list would be to replace all links with bundled references, including the index layers.
However, recall that range queries are the only operations that utilize bundles and only require them as they traverse the range of interest.
As an optimization, we therefore only bundle references at the bottom-most layer (data layer), leaving the index layers as is for use during traversals.

Second, because update operations manipulate multiple links per node, they are linearized using logical flags.
Specifically, insert operations set a \texttt{fullyLinked} flag in the new node after the links of all its \texttt{pred} nodes are updated to point to it. Setting this flag is the linearization point of insert operations. Thus, it is book-ended by the preparation and finalization of the bundles for the predecessor and the new node, similarly to the bundled linked list, by using Algorithm~\ref{algo:linupdateop}. 

Remove operations are handled as follows.
Upon a successful removal, the logical deletion flag is set to linearize the operation, and the bundle entry of the node immediately preceding the target in the data layer is updated.
Then, the references of the predecessor nodes in the index and data layer 
are modified to physically remove the node.

To support linearizable range queries, the skip list defines the two required functions as follows.
\texttt{GetFirstNodeInRange} leverages the traversal over the index layer to find a node in the data layer that points to the first node in the range.
Then it scans to enter the range using the bundles.
\texttt{GetNext} is then used to scan the bottom list, using bundles, and collect the range query's result set, as described in Section~\ref{sec:rangequeries}.
\section{Bundled Binary Search Tree}
\label{sec:citrus}

\begin{figure*}[t]
    \centering
    \begin{subfigure}{\textwidth}
        \centering
        \includegraphics[width=.45\textwidth]{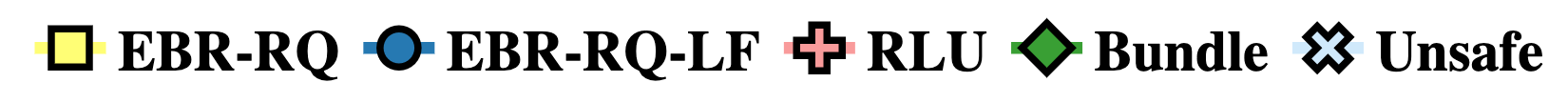}
    \end{subfigure}
    \begin{subfigure}{0.19\textwidth}
        \includegraphics[width=\textwidth]{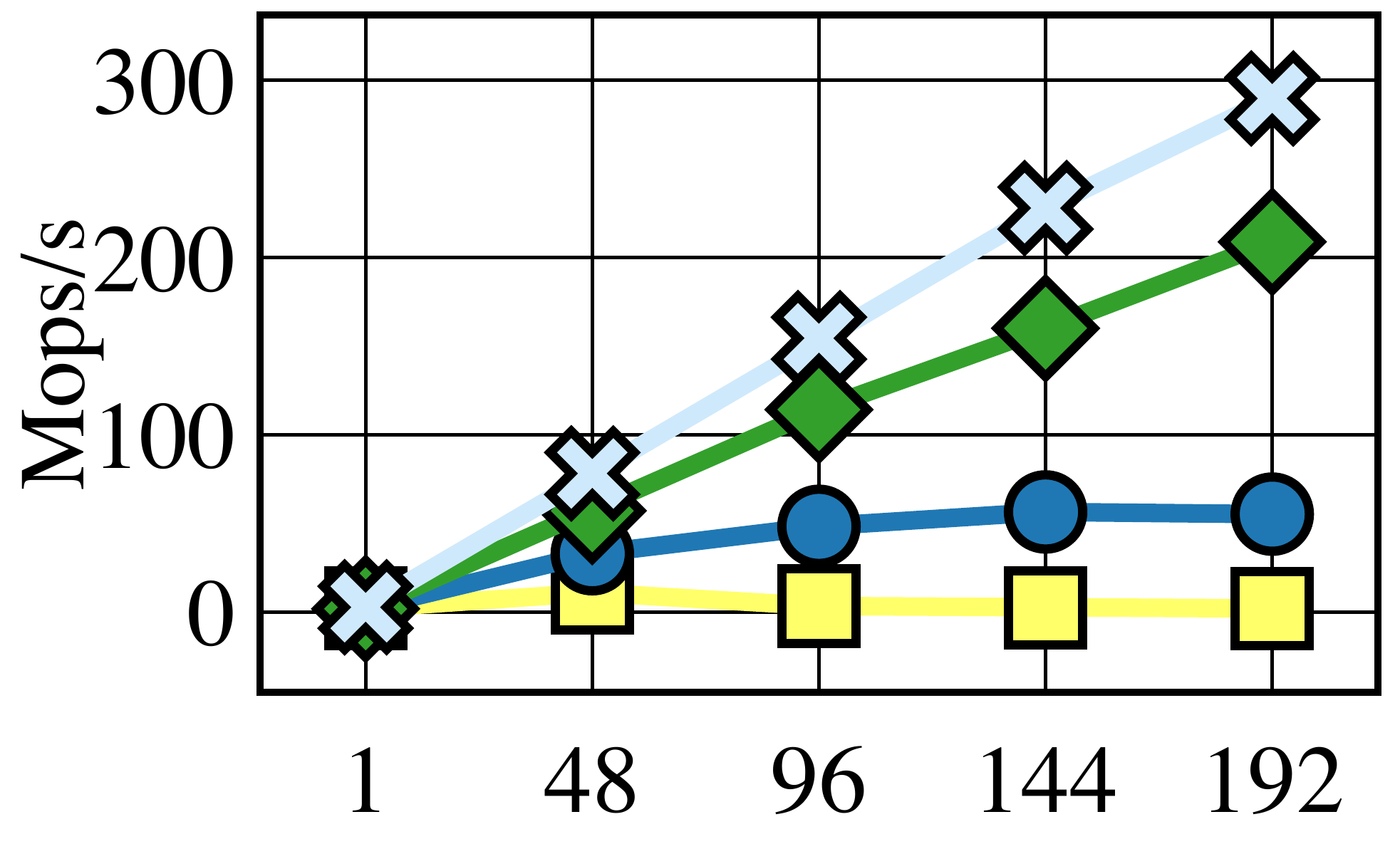}
        \caption{SL, $2-88-10$}
        \label{fig2:a}
    \end{subfigure}
    \begin{subfigure}{0.2\textwidth}
        \includegraphics[width=\textwidth]{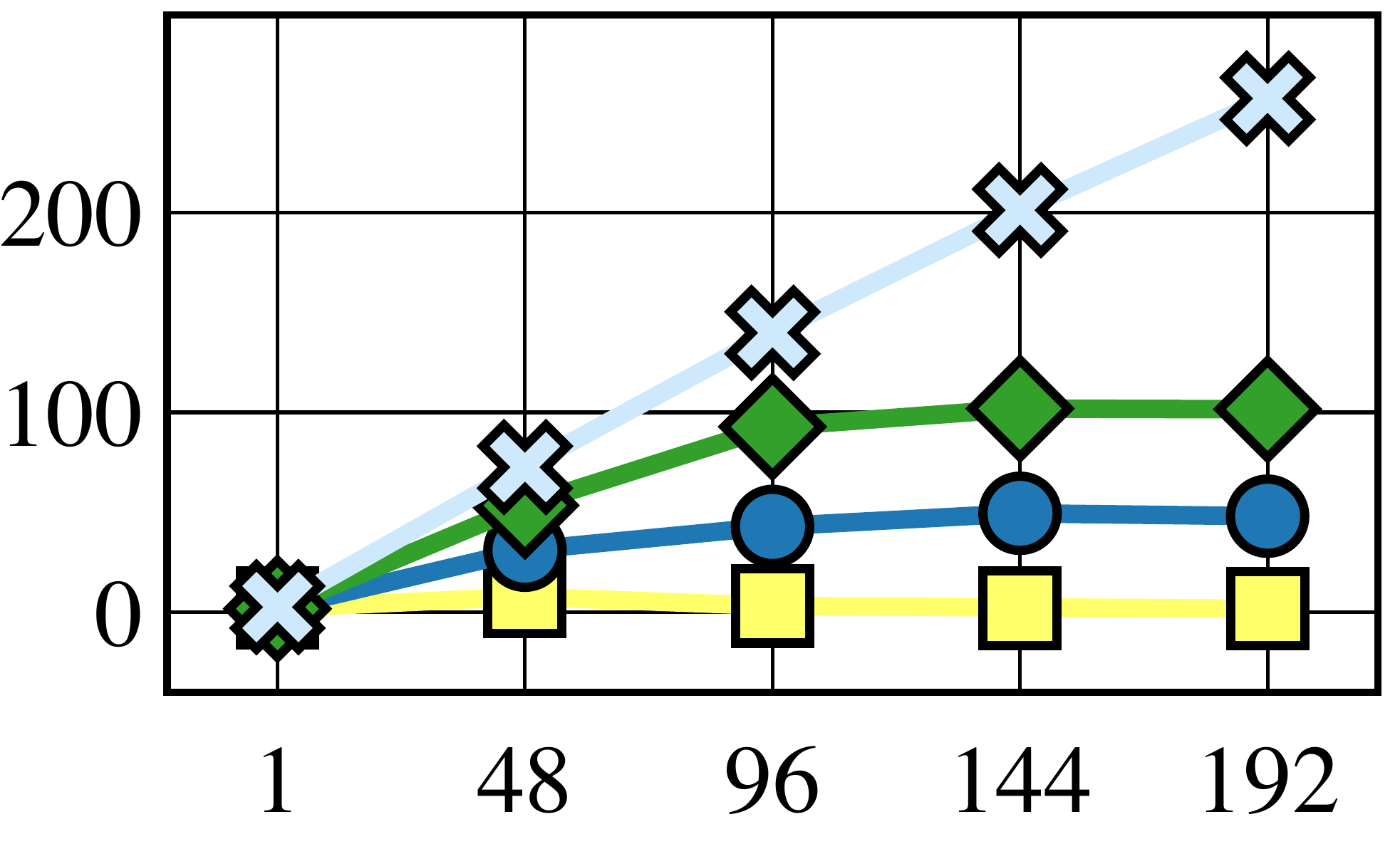}
        \caption{SL, $10-80-10$}
        \label{fig2:b}
    \end{subfigure}
    \begin{subfigure}{0.19\textwidth}
        \includegraphics[width=\textwidth]{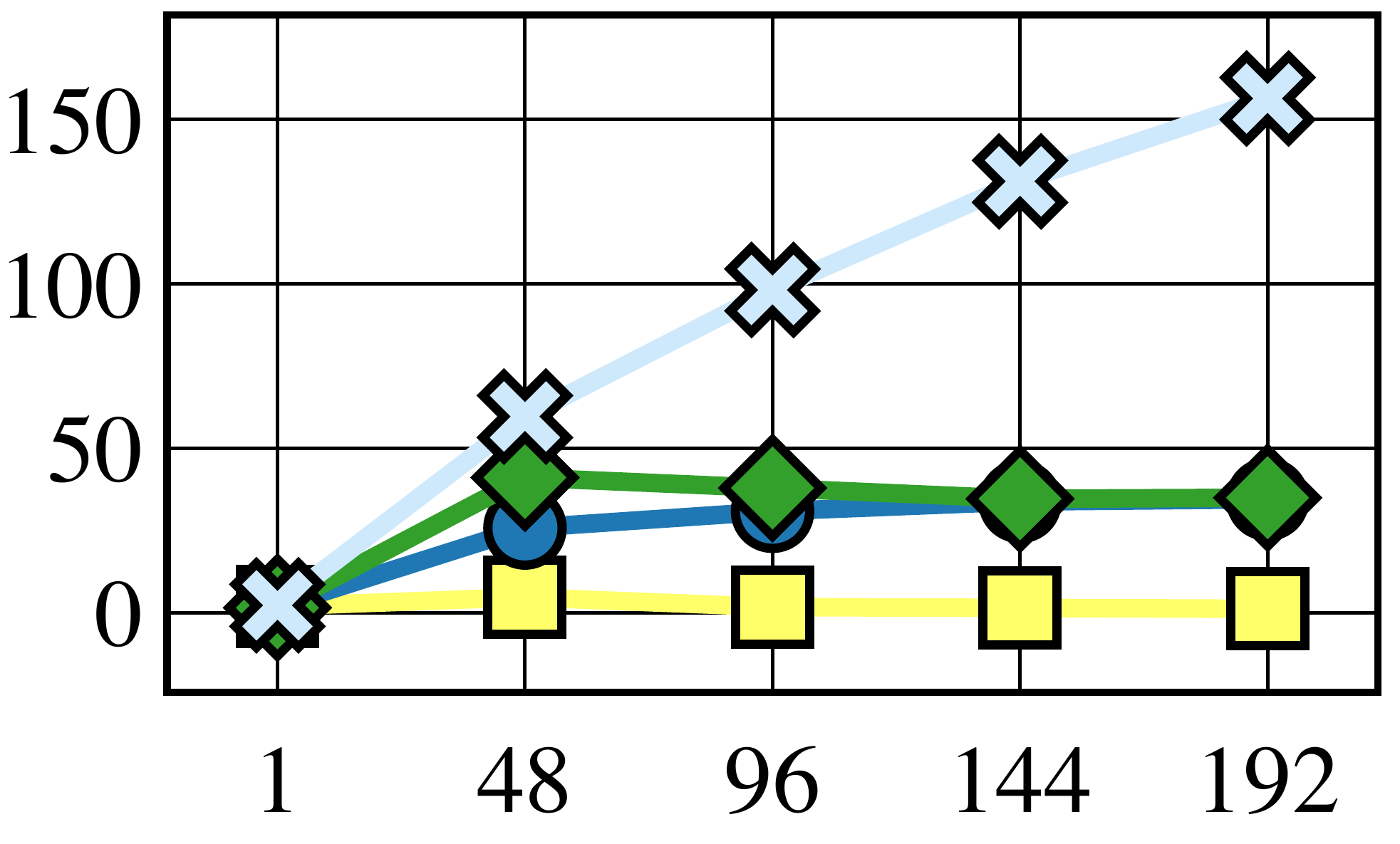}
        \caption{SL, $50-40-10$}
        \label{fig2:c}
    \end{subfigure}
    \begin{subfigure}{0.19\textwidth}
        \includegraphics[width=\textwidth]{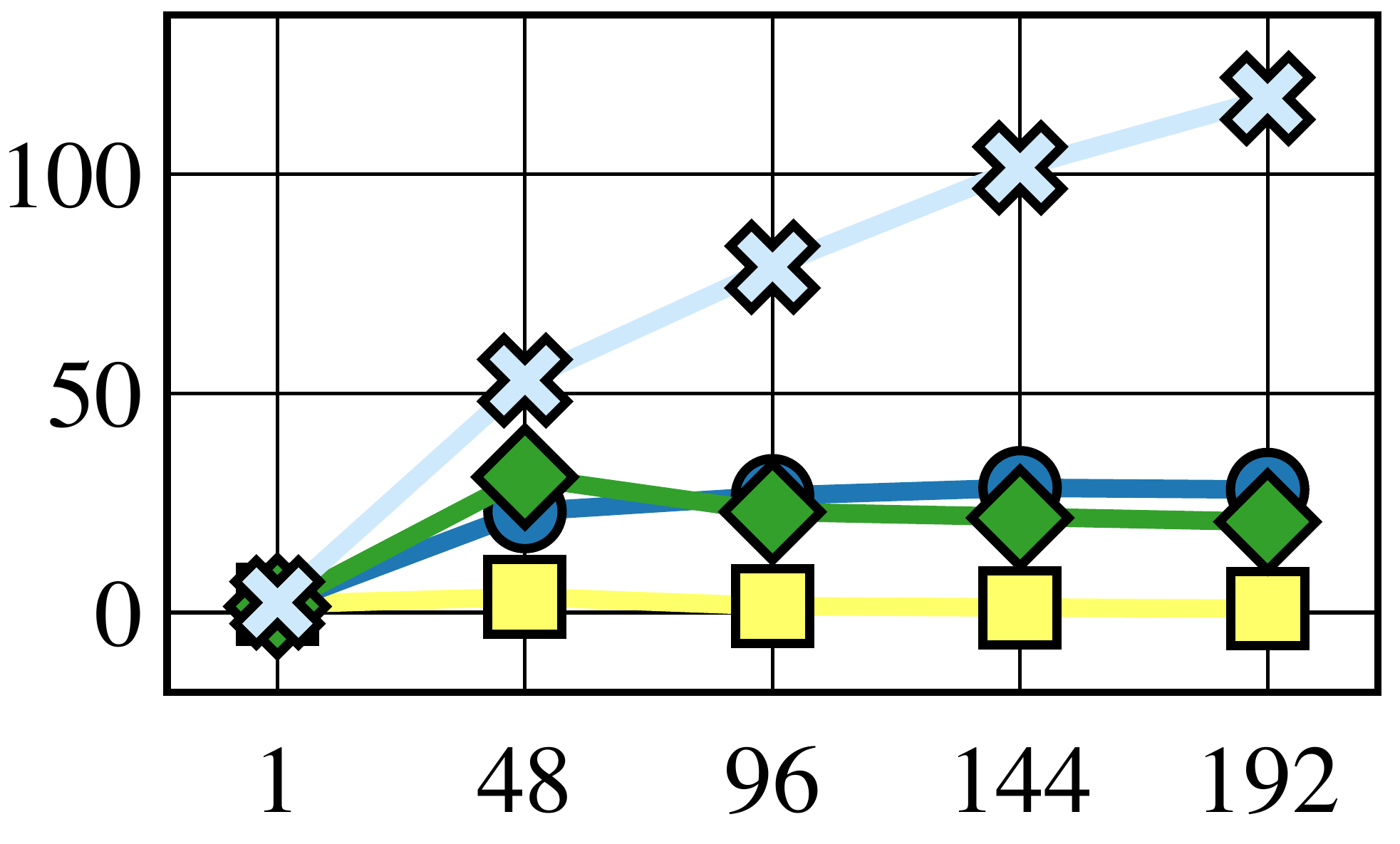}
        \caption{SL, $90-0-10$}
        \label{fig2:d}
    \end{subfigure}
    \begin{subfigure}{0.19\textwidth}
        \includegraphics[width=\textwidth]{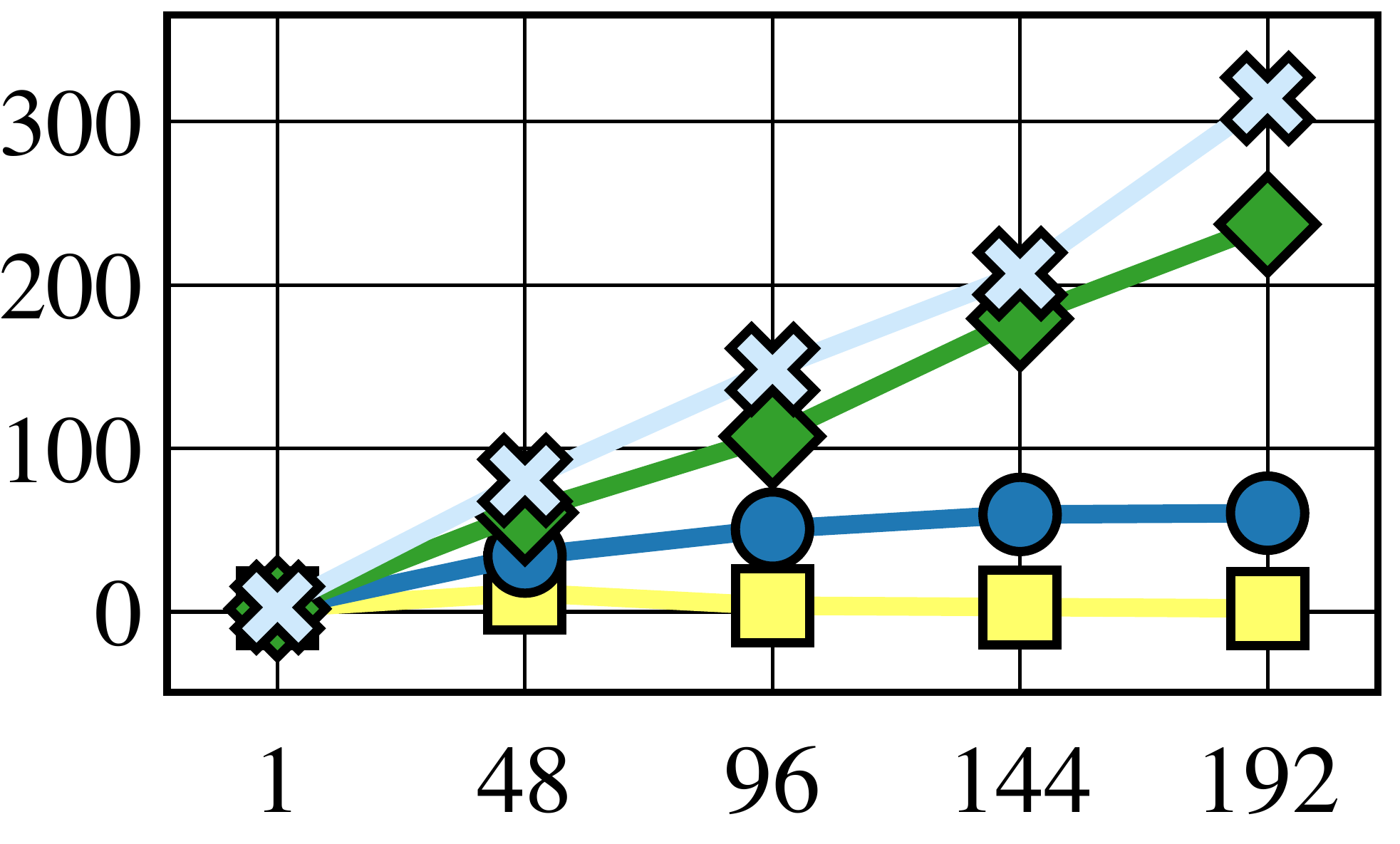}
        \caption{SL, $0-90-10$}
        \label{fig2:e}
    \end{subfigure}\\
    \begin{subfigure}{0.19\textwidth}
        \includegraphics[width=\textwidth]{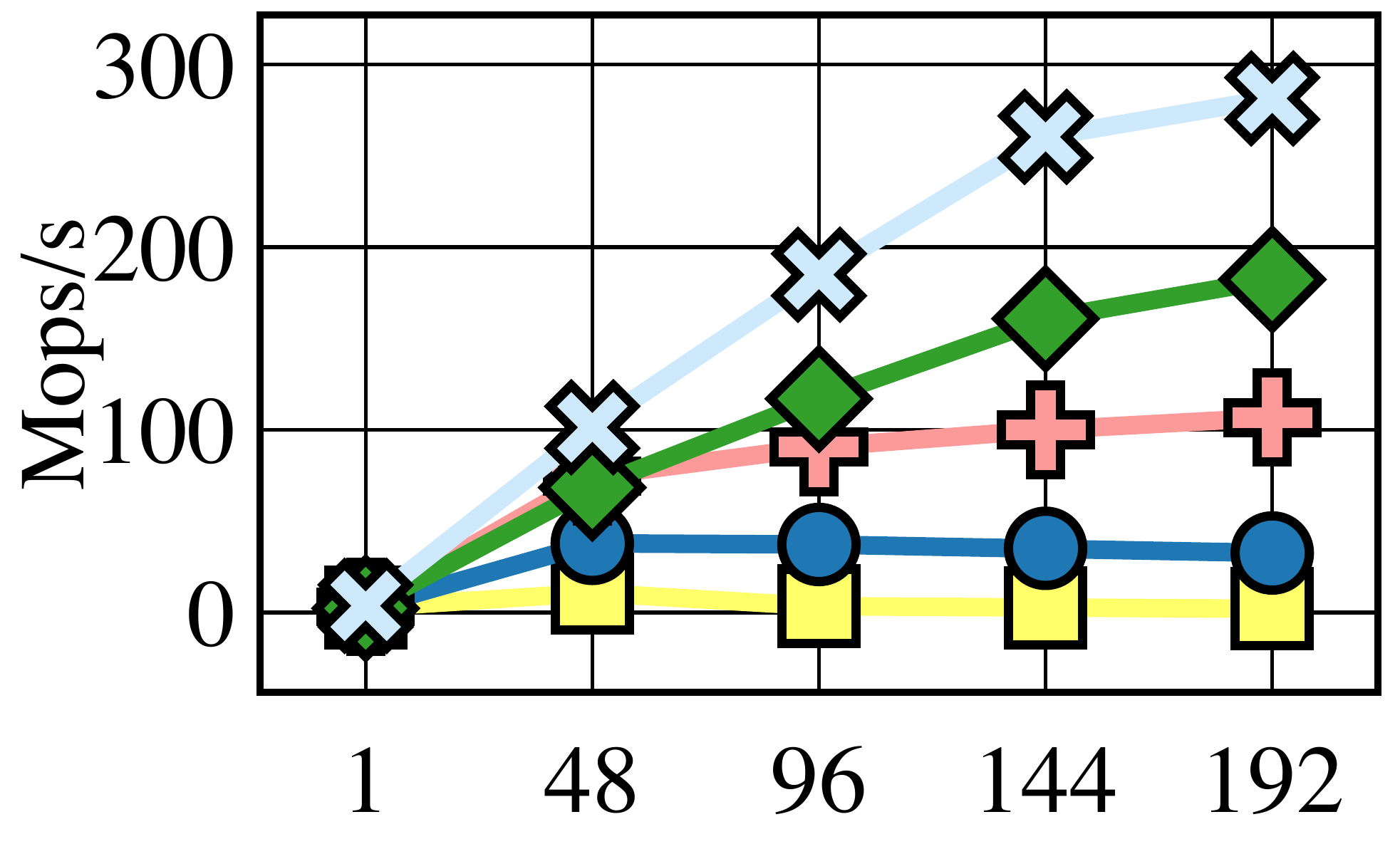}
        \caption{CT, $2-88-10$}
        \label{fig2:f}
    \end{subfigure}
    \begin{subfigure}{0.19\textwidth}
        \includegraphics[width=\textwidth]{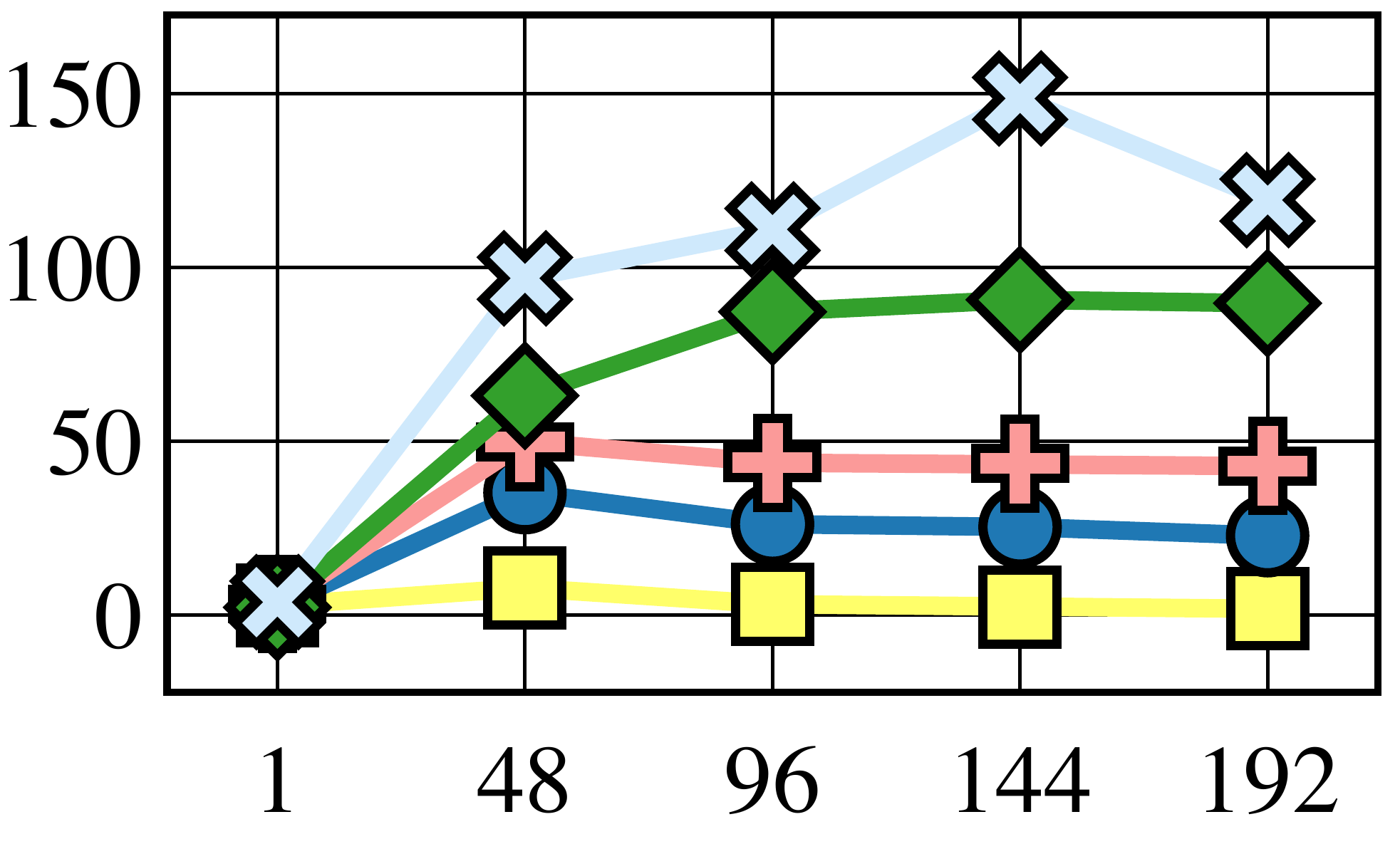}
        \caption{CT, $10-80-10$}
        \label{fig2:g}
    \end{subfigure}
    \begin{subfigure}{0.19\textwidth}
        \includegraphics[width=\textwidth]{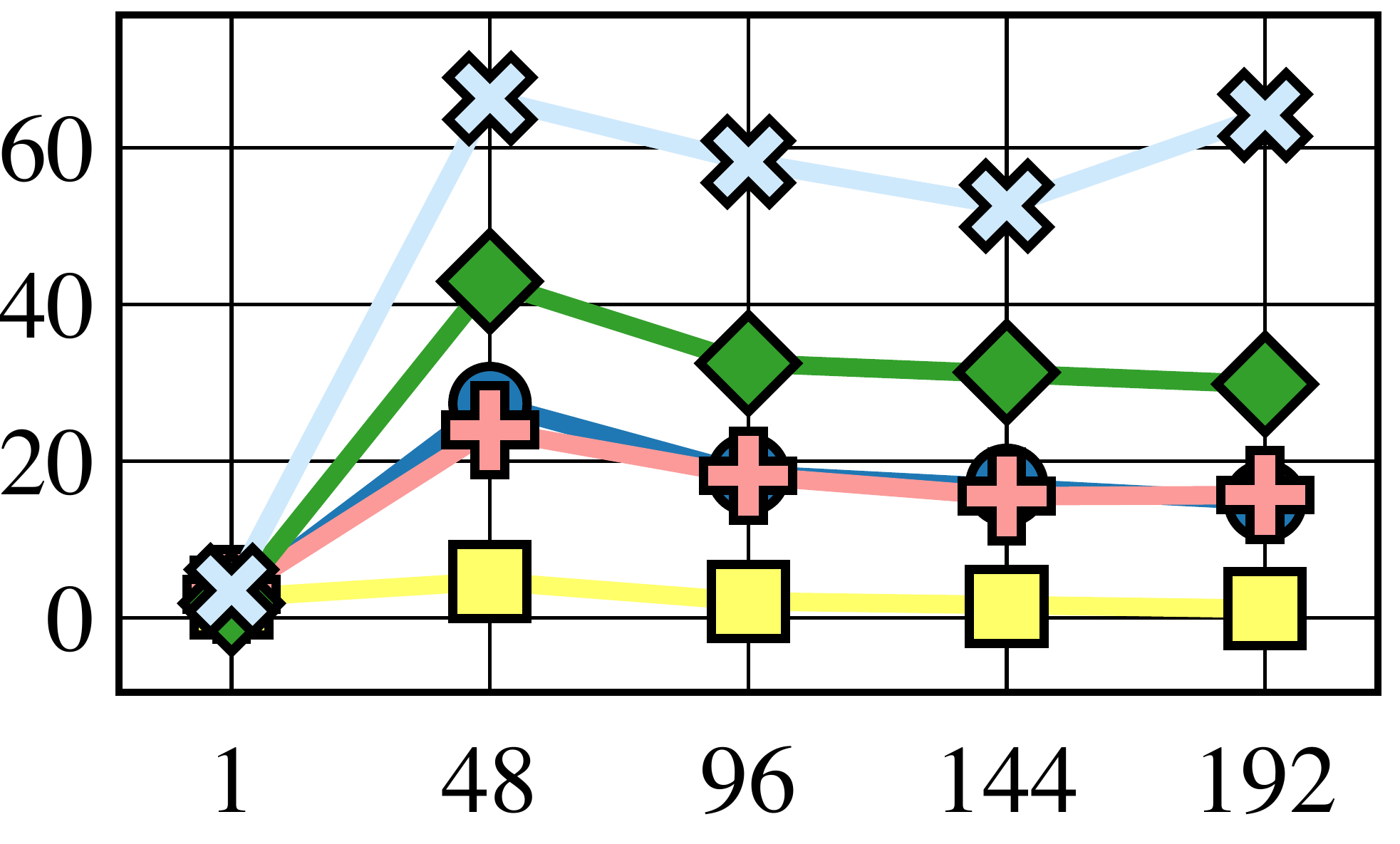}
        \caption{CT, $50-40-10$}
        \label{fig2:h}
    \end{subfigure}
    \begin{subfigure}{0.19\textwidth}
        \includegraphics[width=\textwidth]{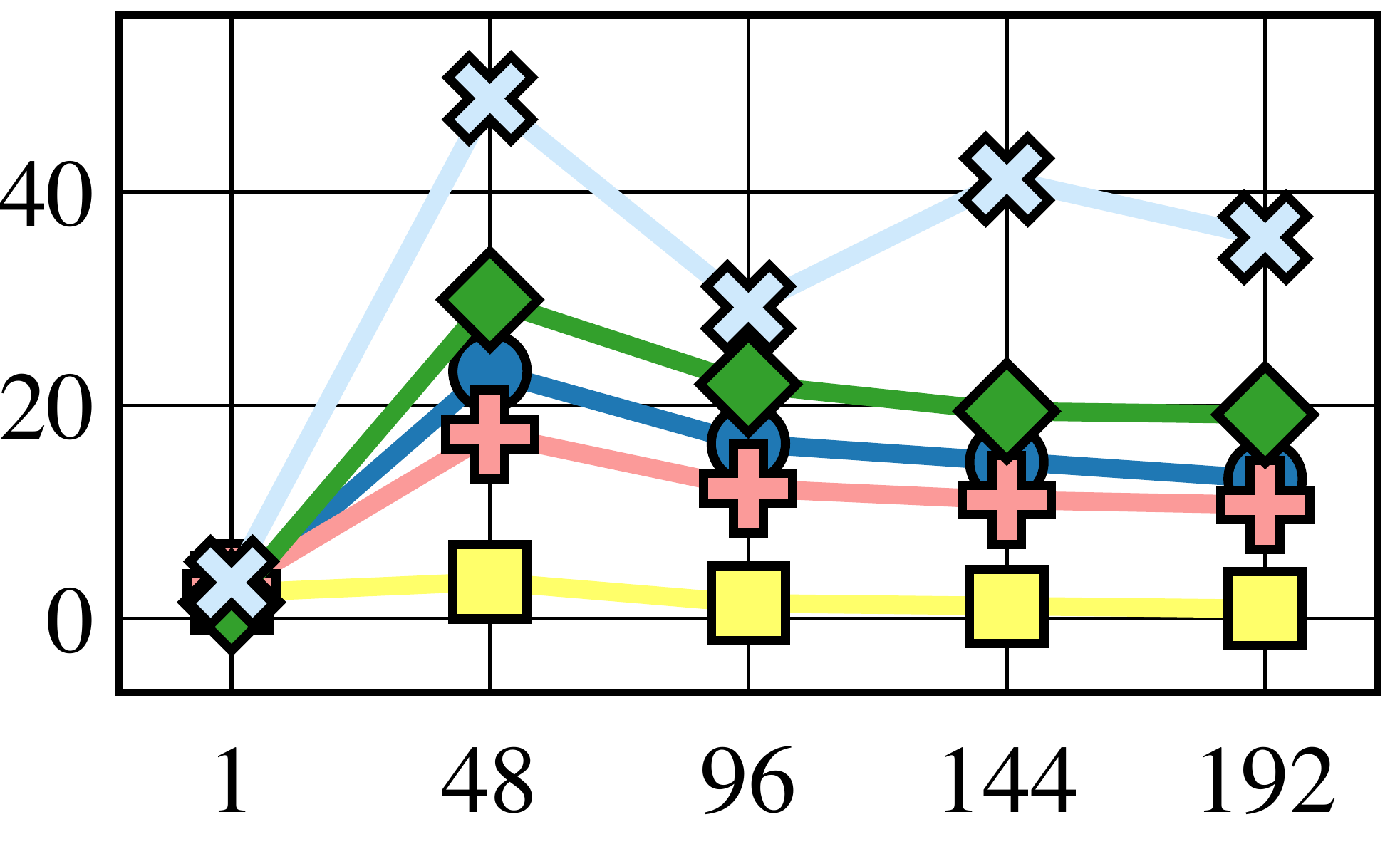}    
        \caption{CT, $90-0-10$}
        \label{fig2:i}
    \end{subfigure}
    \begin{subfigure}{0.19\textwidth}
        \includegraphics[width=\textwidth]{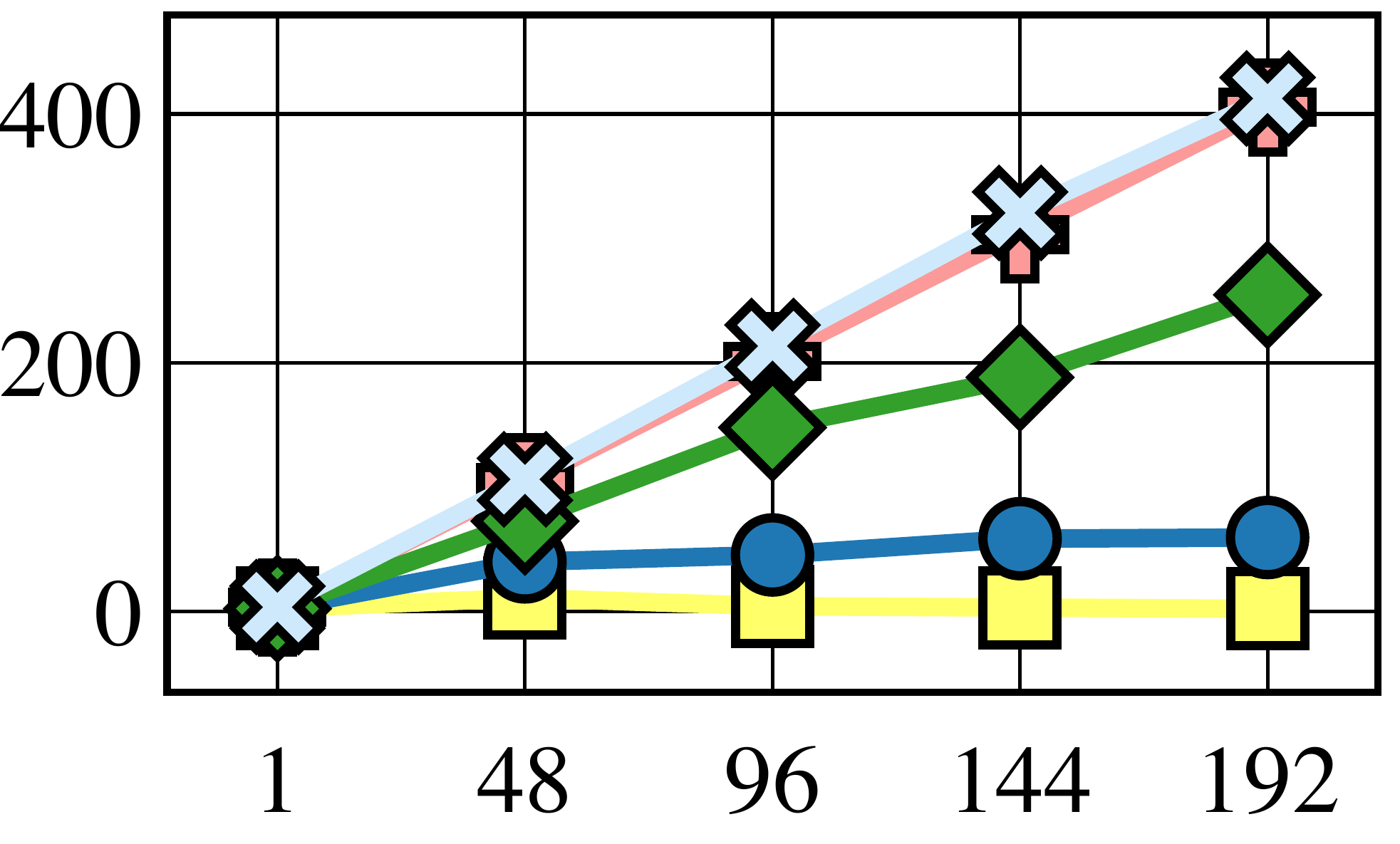}
        \caption{CT, $0-90-10$}
        \label{fig2:j}
    \end{subfigure}
    \caption{Throughput (Mops/s) under various workload configurations for the skip list (SL) and Citrus tree (CT), with the number of threads on the x-axis. Workloads are written as $U-C-RQ$, corresponding to the percentages of update ($U$), contains ($C$) and range queries ($RQ$).}
    \label{fig:fig2}
\end{figure*}

For our bundled tree, we reference the Citrus unbalanced binary search tree~\cite{citrus}, which leverages RCU and lazy fine-grained locking to synchronize update operations while supporting wait-free traversals.
We modify it by replacing each child link of the search tree with a bundled reference.

Citrus implements a traversal enclosed in a critical section protected by RCU's read lock, the required calls of which are wait-free.
This protects concurrent updates from overwriting nodes required by the traversal.
After the traversal, if the current node matches the target, the operation returns a reference to it (named \texttt{curr}), a reference to 
its parent (named \texttt{pred})
and the direction of the child from \texttt{pred}.
Otherwise, the node is not found and the return value of \texttt{curr} is null.
Contains operations simply invoke this traversal then return whether \texttt{curr}'s value is non-null.

Because the Citrus tree is unbalanced, insertions are straightforward and always insert a leaf node. 
Otherwise adhering to the original tree algorithm, insertions are linearized by first preparing the bundle of \texttt{pred} corresponding to the direction of the new node; then by setting the appropriate child, incrementing the global timestamp, and finalizing the bundles.
Lastly, the insert unlocks \texttt{pred} and returns.

The more interesting case is a remove operation, which should address three cases, assuming that the target node \texttt{curr} is found and will be removed.
In the first case, \texttt{curr} has no children and the child of \texttt{pred} pointing to \texttt{curr} is updated along with its bundle.
In the second case, the node to be removed has a child, but only one.
In this scenario, the only child of \texttt{curr} replaces \texttt{curr} as the child of \texttt{pred}.
Again, the bundle corresponding to \texttt{pred}'s child is also updated accordingly.
The last, and more subtle, case is when \texttt{curr} has two children, in which we should replace the removed node with its successor (the left-most node in its right subtree).

In this last case,
both the \texttt{curr}'s successor and its parent are locked. 
Then, following RCU's methodology, a copy of the successor node is created and initialized in a locked state with its children set to \texttt{curr}'s children.
The effect of this behavior
is that possibly four bundles must be modified to reflect the new physical state after the operation take effect.
First, \texttt{pred}'s left or right bundle are modified with an entry referencing the copy of \texttt{curr}'s successor.
Next, both bundles in the copy are also set to \texttt{curr}'s children.
Finally, if the parent of \texttt{curr}'s successor is not \texttt{curr} then its bundle is updated to be null, as the successor is being moved.

In all cases, the remove operation is linearized at the moment the child in \texttt{pred} is changed, making the update visible, and bundles are adjusted along with this linearization point.

Range queries slightly differ from the previous two implementations.
For trees, unlike lists, the node preceding the range (found by \texttt{GetFirstNodeInRange}) is not necessarily a node whose key is lower than the lower bound of the range.
Instead, it is the first node discovered through a depth-first traversal whose child is in the range.
This child is the root of the sub-tree that includes all nodes belonging to the range.
Similar to before, the node reached by the optimistic traversal may not be the correct entry point to the range, and subsequent traversal using bundles may be needed.

Traversing the range follows a depth-first traversal using \texttt{GetNext}.
We keep a stack of nodes to help traverse the subtree rooted at the node returned by \texttt{GetFirstNodeInRange}.
The stack is initialized with the first node in the range.
\texttt{GetNext} pops a node and checks whether its key is lower than, within, or greater than the range.
Next, it adds the node's corresponding children to the stack according to this check.
Finally, if the node is within the range it returns its value to be added to the result set.
Otherwise, it pops another node and performs the above procedure again.

\section{Memory Reclamation}
\label{sec:memreclamation}

%
%
%
%

%

%
%

%

%

%
We rely on EBR to cleanup both physically removed nodes and no longer needed bundle entries because, as already assessed by~\cite{ebr-rq}, quiescent state memory reclamation~\cite{rcu} (a generalized form of EBR) mirrors the need for a range query to observe a snapshot of the data structure.
A complete discussion of the details regarding memory reclamation can be found in the supplemental material.
Although the experiments in Section~\ref{sec:evaluation} were performed without enabling memory reclamation, the same conclusions are drawn with respect to competitors when memory is reclaimed.

\begin{figure*}[t]
    \centering
    \begin{subfigure}{.65\textwidth}
        \centering
        \includegraphics[width=\textwidth]{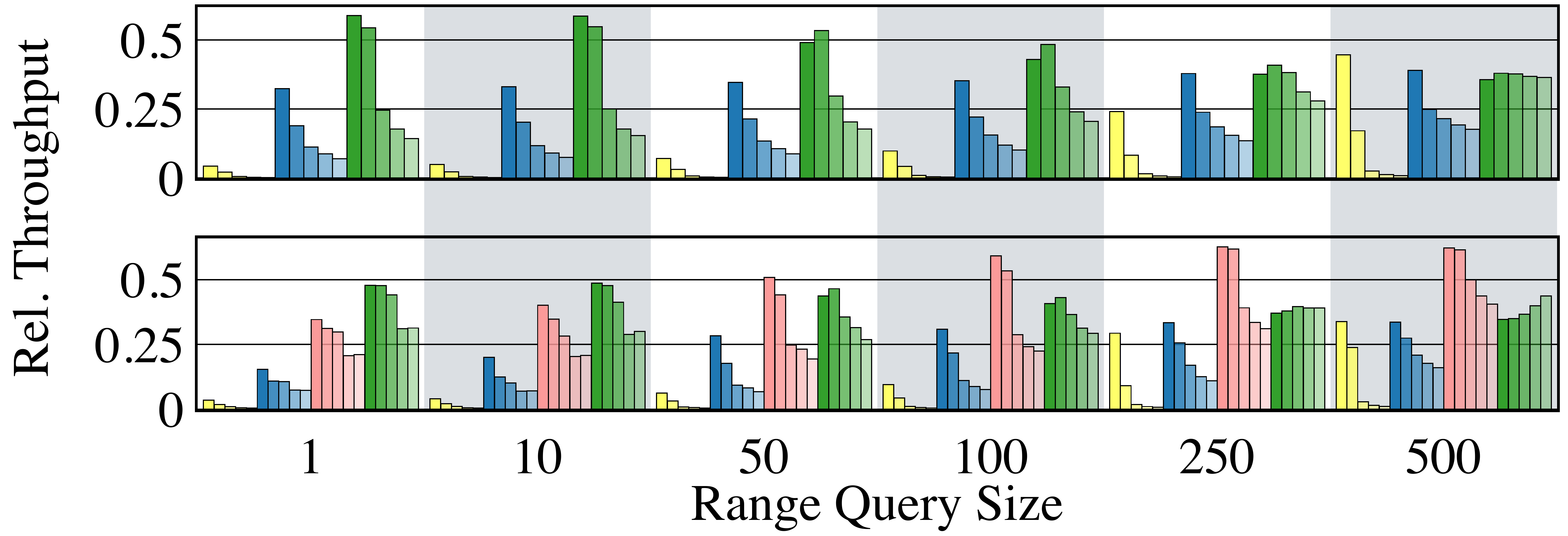}
    \end{subfigure}
    \begin{subfigure}{.29\textwidth}
        \centering
        \includegraphics[width=\textwidth]{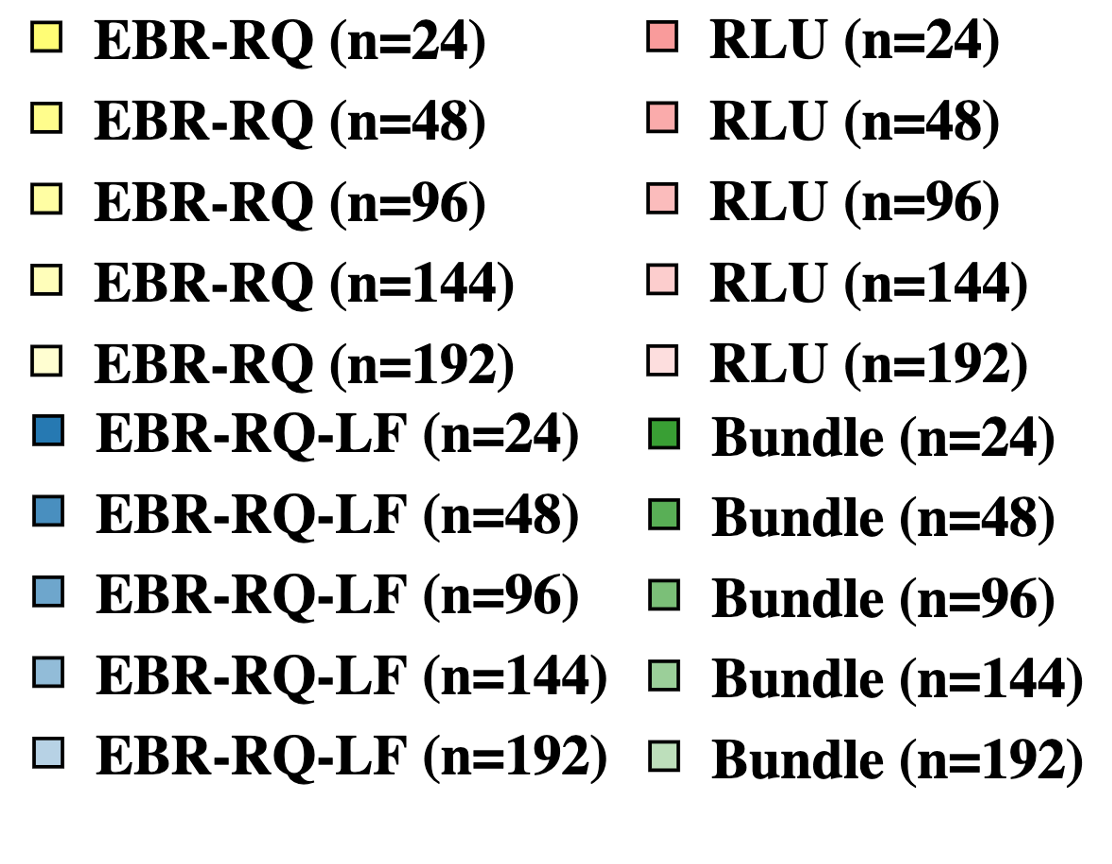}
    \end{subfigure}
    \caption{Throughput, relative to Unsafe, for different range query lengths for skip list (top) and Citrus tree (bottom) with a $50-0-50$ workload. Results at each cluster are first organized by competitor (in the following order: EBR-RQ, EBR-RQ-LF, RLU, Bundle), then ordered by thread count $n$; the right-most bar of each group being the highest thread count.}
    \label{fig:rqlen}
\end{figure*}
\section{Evaluation}
\label{sec:evaluation}

In each of the following experiments we compare our approach (named \textit{Bundle} hereafter) with \textit{RLU}~\cite{rlu} and two variants of Arbel-Raviv and Brown's technique based on epoch-based reclamation~\cite{ebr-rq} (referred to as \textit{EBR-RQ} and \textit{EBR-RQ-LF} hereafter).
EBR-RQ uses a readers-writer lock to protect its global epoch counter and EBR-RQ-LF is lock-free.
The lock-free version still locks data structure nodes, but the infrastructure supporting linearizable range queries is lock-free.
Note that our technique uses EBR but strictly for memory management; whereas, EBR-RQ and EBR-RQ-LF rely on EBR's internals in order to support linearizable range queries (see Section~\ref{sec:rel-work}).

As a reference for performance we implement \textit{Unsafe}, a version of each data structure whose range queries traverse without performing any consistency checks, while still providing linearizable primitive operations.
For readability, we do not include the performance of Snapcollector because its throughput is significantly
slower than all other competitors.

We integrate our data structures into an existing framework~\cite{ebr-rq} to develop and benchmark a variety of data structures supporting linearizable range queries, including RLU-base linked list and Citrus tree, and the EBR-base variants of all three data structures.
The code is written in C++ and compiled with \texttt{-std=c++11 -O3 -mcx16}.
All tests are performed on a machine equipped with four Xeon Platinum 8160 processors, for a total of 96 physical cores and 192 hyper-threaded cores, running Red Hat Enterprise Linux 7.6.

\subsection{Bundled Data Structure Performance}
\label{sec:microbench}

For each of the following experiments the data structure is first initialized with half of the keys in the key range; all updates are evenly split between inserts and removes to ensure size stability.
Threads execute a given mix of update, contains, and range query operations. 
Workloads are reported as $U-C-RQ$, where $U$ is the percentage of updates, $C$ is the percentage of contains and $RQ$ is the percentage of range queries.
Target keys are procured uniformly. 
All reported results are an average of three runs of three seconds each, except where noted.
The key range of each data structure is as follows: the lazy list is 10,000 and the skip list and BST are both 100,000.

\textbf{Varying Workload Mix.}
A side effect of logarithmic traversals in the skip list and Citrus tree is that the costs of supporting linearizable range queries is more visible compared to the linked list, which have linear asymptotic bounds.
Thus, we defer an analysis of our bundled linked list until later.
We report the operation throughput of different workload mixes as a function of thread count.
The range query percentage is fixed at 10\%, while varying the update and contains percentages.
We plot the total throughput for both the skip list and Citrus tree in Figures~\ref{fig2:a}-\ref{fig2:e} and Figures~\ref{fig2:f}-\ref{fig2:j}, respectively.

Our first general observation is that our bundled data structure outperform all linearizabe competitors when the workload is mixed (the first three columns of Figure~\ref{fig:fig2}).
These three configurations represent a wide class of workloads, contrasted with the two right-most columns that represent corner case workloads and are discussed separately.
Under mixed loads, Bundle achieves maximum speedups over the nearest competitor of 3.9x (skip list, Figure~\ref{fig2:a}) and 2.1x (Citrus tree, Figure~\ref{fig2:g}).
Both maximums occur when the workload is dominated by reads and occur at the highest number of threads tested.

The above behavior is the result of two design characteristics of bundling. 
First, single element contains are not instrumented in any way.
Second, range queries only wait for ongoing updates that are localized in the target key range.
In low update percentage workloads, this provides bundling with the advantage.
Both EBR-RQ and EBR-RQ-LF incur significant overhead in this particular configuration.
The former due contention on a global lock; the latter due to the use of a costly double-compare single-swap primitive (DCSS), which impacts both range queries and contains operations.

The performance gap between Bundle and its competitors narrows as the percentage of updates increases to 50\% and 90\%.
While RLU is faced with additional dereference logic for reads, its primary bottleneck lies in the synchronization step required by writes waiting for ongoing reads to finish.
This behavior leads to poor performance in update-intensive workloads.
EBR-RQ and EBR-RQ-LF perform better relative to RLU under these circumstances and barely outperforms bundling in a 90\% update workload on the skip list.
In bundling, the primary source of overhead is updates contending on an atomically incremented global timestamp and temporarily stalling range queries.
Regardless, it performs comparably or better in all but one of these configurations.

To better understand the cases in which Bundle performance is surpassed, we note that RLU and EBR-RQ prefer workloads at opposing ends of the configurations spectrum.
In the read-only setting (Figure~\ref{fig2:e} and~\ref{fig2:j}), RLU performs well in contrast with EBR-RQ and EBR-RQ-LF.
Of particular note, RLU achieves performance nearly equivalent to Unsafe in the Citrus Tree.
In the absence of updates, RLU's execution pays little cost for reads and range queries.
However, recall that even a low percentage of updates is enough to cause this impressive performance to collapse (see Figure~\ref{fig2:f}), namely from the synchronization enforced by writers (i.e., RLU-sync).
Hence, when a workload is primarily updates (Figure~\ref{fig2:d} and~\ref{fig2:i}), RLU incurs even higher overhead.
On the other hand, EBR-RQ and EBR-RQ-LF increment a global epoch counter and validate their snapshot, which leads to the degradation of performance in read-only workloads to be more than for update-intensive ones.

Bundling manages the trade-off between update-intensive and read-only workloads effectively.
The overhead of updating bundles is relatively low, and is fine-grained, which improves upon RLU's synchronization. Only traversing the necessary nodes improves upon both EBR-RQ and EBR-RQ-LF. 
Hence, performance stability across different workloads is an important byproduct of bundled data structures.

Unlike RLU and the EBR-based techniques, bundling does not concentrate overhead, but distributes the responsibility of linearization between updates and range queries.

\textbf{Varying Range Query Size.}
~Figures~\ref{fig:rqlen} shows the relative throughput over Unsafe for a skip list (top) and Citrus tree (bottom) when performing equal percentages of updates and range queries at increasing range query sizes (from 1 to 500).
The workload roughly corresponds to the middle column in Figure~\ref{fig:fig2}, having a $50-0-50$ mix with the intention of avoiding bias toward either competitor.

Under the given workload, bundling outperforms all linearizable competitors at large numbers of threads, regardless of range query size.
For all numbers of threads, we outperform EBR-RQ and EBR-RQ-LF. In fact, regardless of the length of the range query EBR-RQ-LF, on average, checks an additional 300 nodes in the limbo at 96 threads (and 600 nodes at 192 threads), accounting for the majority of its execution time.
Since RLU's synchronization overhead is smaller at fewer threads, the relative cost of traversing bundles is apparent, but only when range queries are long.
Regardless, when the thread count is high, the impact of the use of bundles is less than the synchronization required by RLU's updates and Bundle regains its dominance.

\textbf{Linked Lists.}
Because traversals dominate the runtime of linked lists they provide less insight into algorithm behavior. The linear asymptotic bound causes the best competitors to behave nearly identically to Unsafe. This includes Bundle and the two EBR variants.
The worst competitor (i.e., RLU), on the other hand, has a relative performance of 0.97x ($0-90-10$), 0.87x  ($2-88-10$), 0.70x  ($10-80-10$), 0.42x  ($50-40-10$) and 0.40x  ($90-0-10$) when compared to Unsafe at 96 threads.

\textbf{Weakening Linearizability.} In additional experiments, included in the supplementary material for space constrains, we measured the performance gains for range queries when linearizability is relaxed by only updating the global timestamp every $T$ operations. In majority update workloads this strategy offers 2x better performance when $T=5$ and nearly 3x when $T=50$.

\subsection{Database Integration Performance}

The following results were collected for TPC-C benchmark with 10 warehouses using DBx1000~\cite{stonebreaker}, an in-memory database system, integrated with our bundled skip list and Citrus tree. The data structures implement the database indexes.

Specifically, we use the \texttt{NEW\_ORDER} (50\%), \texttt{PAYMENT} (45\%) and \texttt{DELIVERY} (5\%) transaction profiles.
The \texttt{DELIVERY} profile is particularly interesting since its logic includes a range query over the index representing the new order table, ordered by \texttt{order\_id}, with the goal of selecting the oldest order to be delivered. Next, the order is deleted to prevent subsequent \texttt{DELIVERY} transactions from delivered the same order again. In our experiments, the range query selects the oldest order in the last 100 orders.
A \texttt{PAYMENT} transaction performs a range query on the customer index to look up a customer by name with 60\% probability.
\texttt{NEW\_ORDER} modifies multiple tables and updates their indexes accordingly, including the new order index.

\begin{figure}[h]
    \centering
    \begin{subfigure}{0.4\linewidth}
        \includegraphics[width=\textwidth]{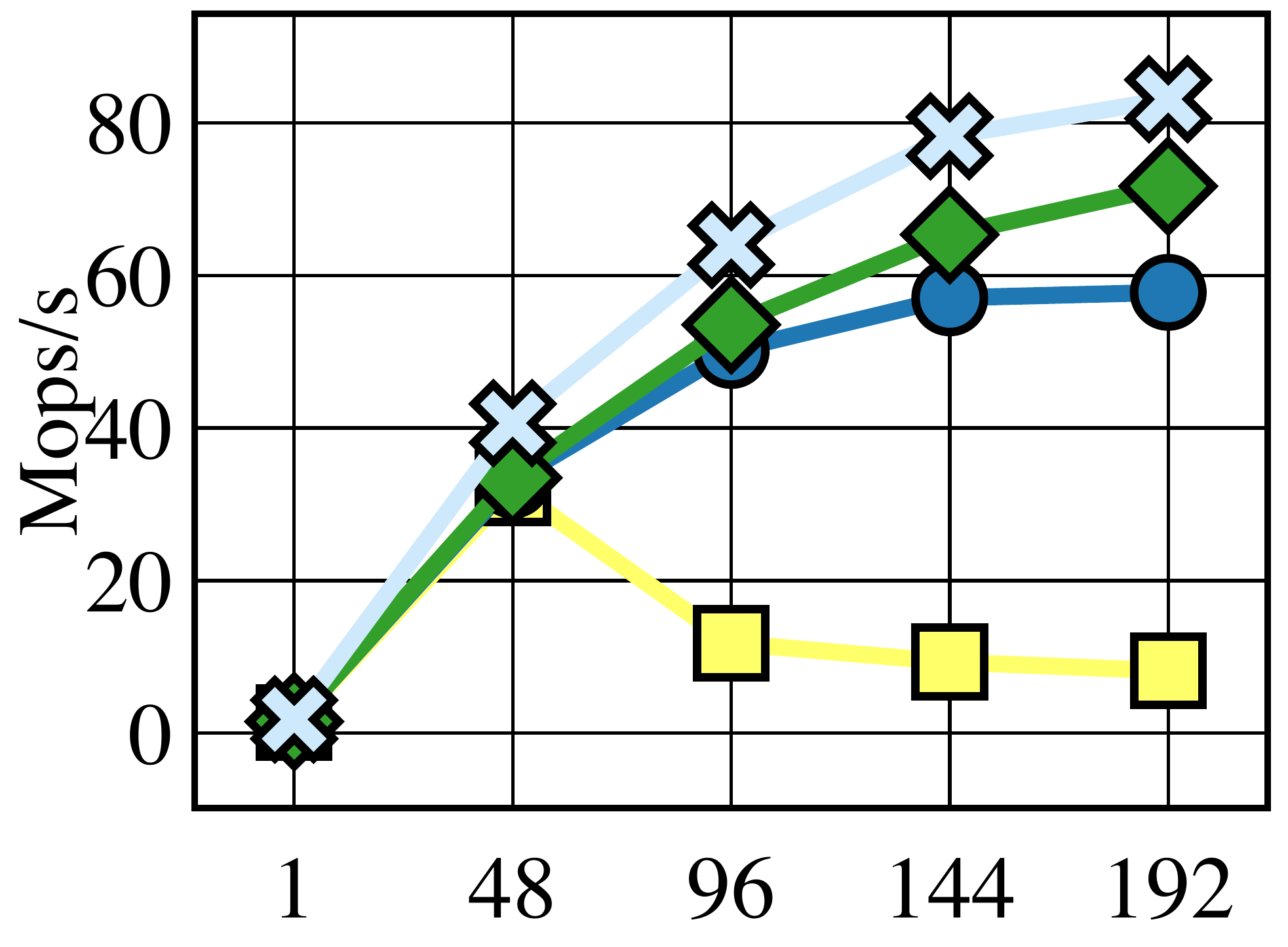}
        \caption{Skip list}
        \label{fig:dbskiplist}
    \end{subfigure}
    \hspace{-5pt}
    \begin{subfigure}{0.4\linewidth}
        \centering
        \includegraphics[width=\textwidth]{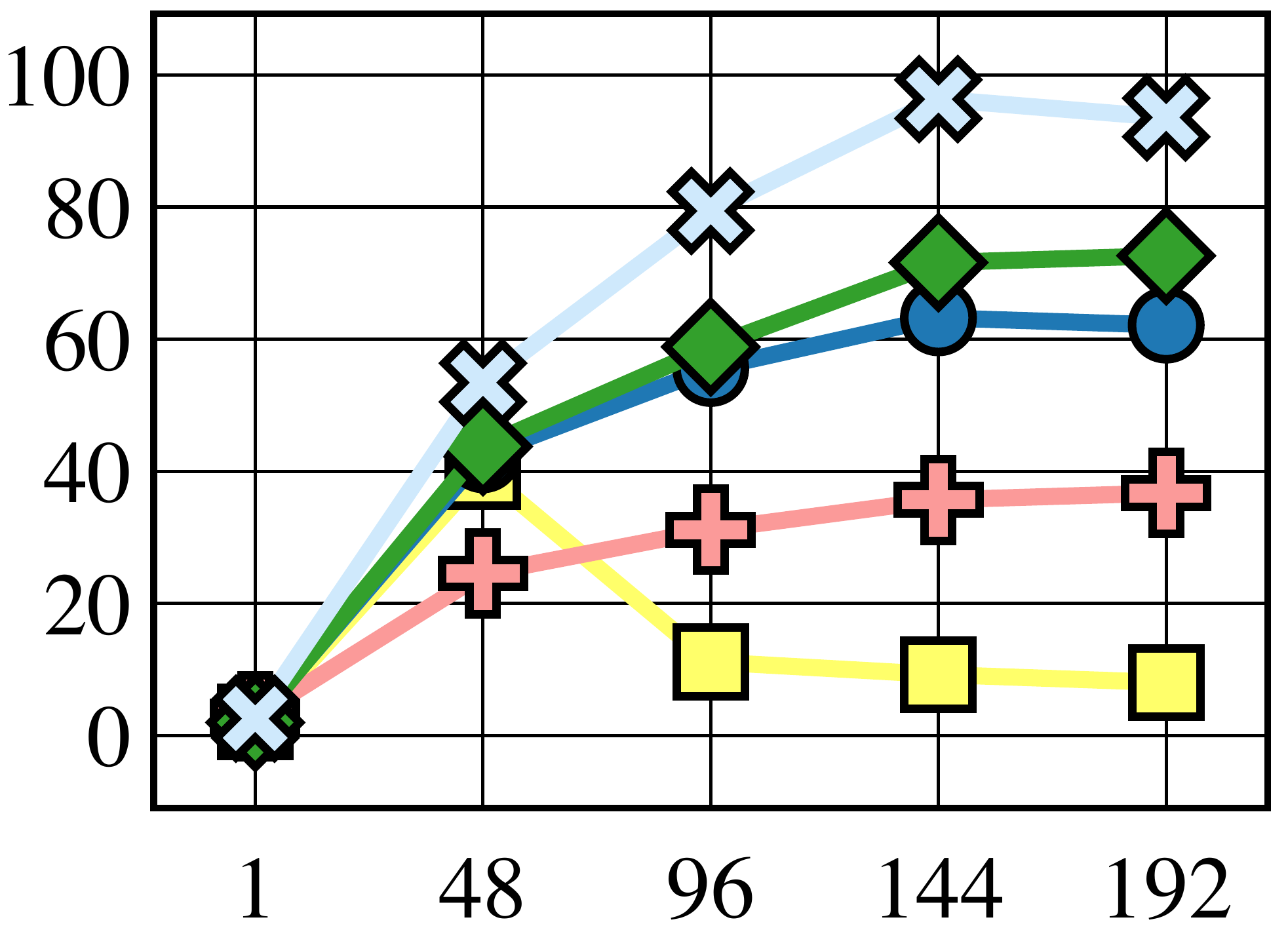}
        \caption{Citrus tree}
        \label{fig:dbtree}
    \end{subfigure}
    \begin{subfigure}{0.19\linewidth}
        \centering
        \includegraphics[width=\textwidth]{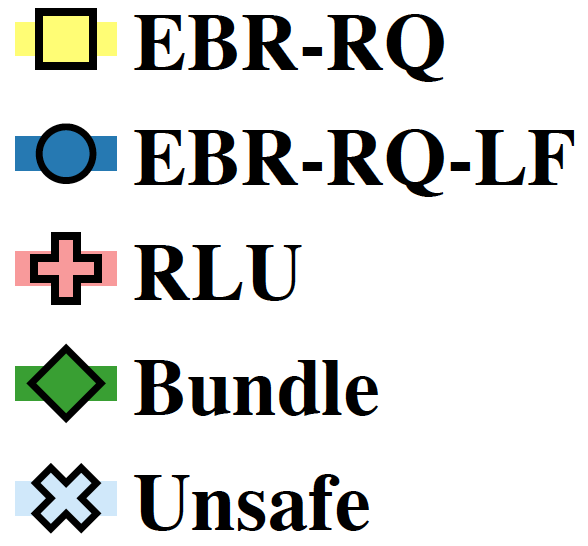}
    \end{subfigure}
    \caption{Throughput (Mops/s) of index operations in DBx1000 running the TPC-C benchmark.}
    \label{fig:macrobench}
\end{figure}

We report the total throughput across all indexes for the skip list and Citrus tree in Figure~\ref{fig:macrobench}.
Note that we elide a comparison against the baseline DBx1000 index since it is a hashmap and does not support range queries.
When isolating performance metrics to indexes only, bundling outperforms all  competitors regardless of the number of threads used. 
In both data structures, EBR-RQ and EBR-RQ-LF follow trends observed previously.
We also measure the overall system throughput (i.e., transactions committed per second), but do not include the plots due to space constraints.
Summarizing the findings, the performance of Bundle over an Unsafe index (non-serializable) is on average 3.6\% worse for the skip list and 12.5\% for the Citrus tree. These results show the effectiveness of bundling when integrated in large systems even under skewed workloads, as is the case for TPC-C.

In the TPC-C application workload, RLU can take advantage of its highly efficient range queries while EBR-RQ and EBR-RQ-LF may benefit from 100\% updates on some indexes.
Unlike its competitors however, Bundle does not sacrifice performance in one case for higher performance in the other, which leads to overall better throughput.
This demonstrates that our more performance-stable design is better suited for systems that have different workloads on different internal data structures, as is typical in database systems, without the need for multiple implementations each targeting specific workload distributions.

\section{Conclusion}
\label{sec:conclusion}

In this paper we presented three concurrent linked data structure implementations 
deploying a novel building block, called bundled references, to enable range query support. Bundling data structure links with our building block shows that the coexistence of range query and update operations does not forgo achieving high-performance.

%

\newpage
\bibliography{references}
\newpage
~
\newpage
\appendix
\section*{Supplementary Material}
\section{Weakening Linearizability}
\label{weakening}

By reducing the frequency with which threads increment the global timestamp, we can measure the impact of contention on the global timestamp required by bundling.
To do so, we record the throughput with threads modifying \texttt{globalTs} only after performing a configurable number of updates $T$.
Similar to the previous experiment, the workload has a $50-0-50$ distribution.

\begin{figure}[h]
    \scriptsize
    \centering
    \begin{subfigure}{\linewidth}
        \centering
        \includegraphics[width=.85\textwidth]{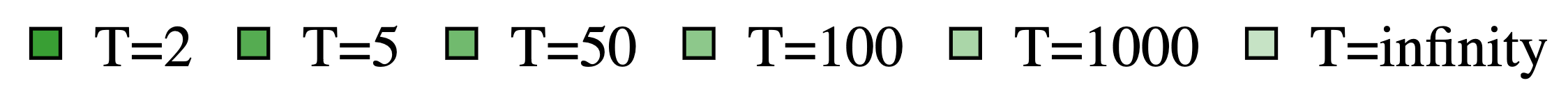}
    \end{subfigure}
    \begin{subfigure}{\linewidth}
        \centering
        \includegraphics[width=.85\textwidth]{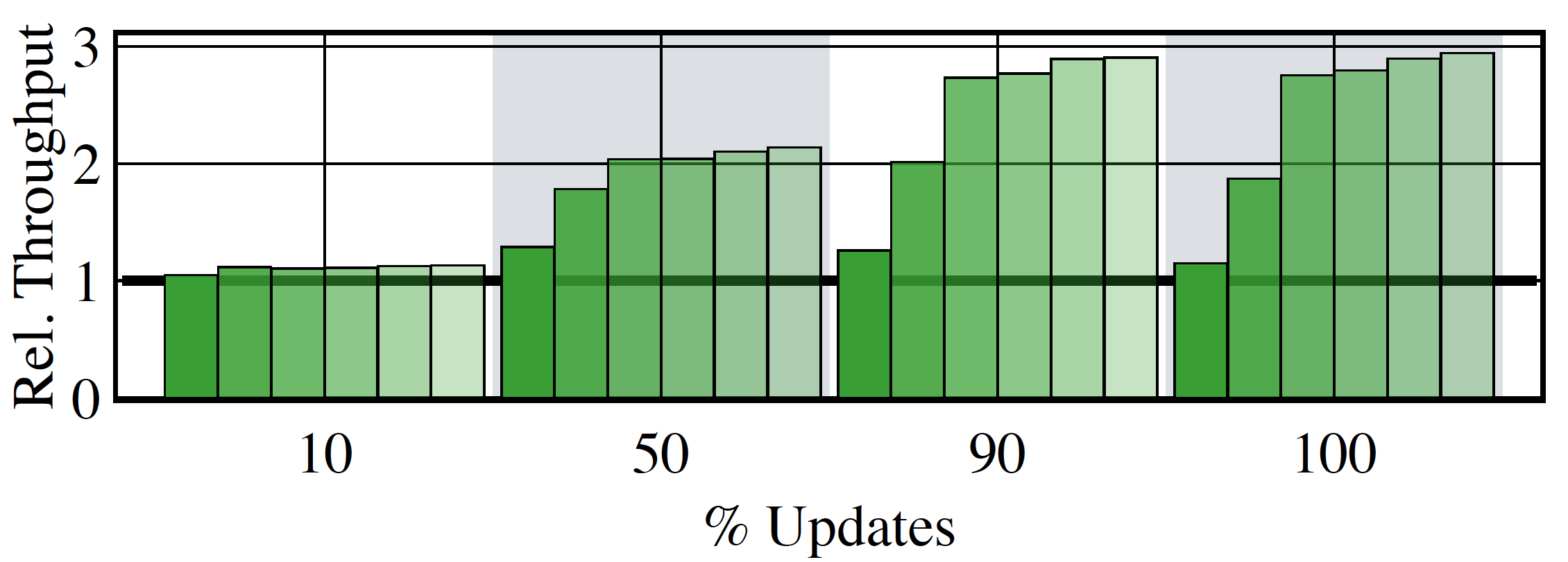}
    \end{subfigure}
    \caption{Throughput relative to the linearizable  implementation of a bundled skip list for various threshold values ($T$), represented by shade, at 96 threads.}
    \label{fig:relax}
\end{figure}

Figure~\ref{fig:relax} shows the results of our experiments for the bundled skip list under different workload distributions at 96 threads.
For read-mostly workloads, changing the threshold $T$ has little benefit since most of the operations do not increment the global timestamp.
At 50\% updates, however, the effects of reducing the frequency of writes to \texttt{globalTs} lead to a 2x improvement when $T=50$.
The improvement decreases at lower thresholds, reaching roughly 25\% when $T=2$.
For write intensive workloads, this improvement is amplified, and close to 3x improvement is observed.
Similar conclusions have been reached for different numbers of threads, as well as for the bundled Citrus tree.

For each group of bars, the right most one illustrates performance when relaxation is at its most extreme (i.e. $T=\infty$).
The effect is an ideal range query that never waits for pending timestamps and always utilize the first entry.
Interestingly, when compared to $T>50$, there is little performance gain for complete relaxation.

Clearly, if update operations do not always increment the global timestamp linearizability is weakened because range queries may be prevented from observing the freshest snapshot of their range.
As a side effect of our design, if an application tolerates such a relaxation, our bundled data structure design allows for a simple and tunable mechanism to adjust the level of freshness of range queries. 
Further investigation about the correctness guarantees of such a weakened relaxed version of our bundled data structures is left as future work.

\section{Memory Reclamation}
\label{appsec:memreclamation}

In this section we show: \textit{i)} how a well-known memory reclamation technique, such as epoch-base memory reclamation (EBR)~\cite{ebr}, can be easily integrated into our bundled data structures to safely manage memory after physically removing nodes; and \textit{ii)} a simple policy (enabled by our design) to recycle no longer needed bundle entries.

We decide to rely on EBR because, as already assessed by~\cite{ebr-rq}, quiescent state memory reclamation~\cite{rcu} (a generalized form of EBR) mirrors the need for a range query to observe a snapshot of the data structure.
In fact, this reclamation technique waits for a grace period to elapse before freeing memory, thus allowing range queries to safely reference nodes removed concurrently.
Specifically, we use a variant of EBR, called DEBRA~\cite{debra}, that stores per-thread limbo lists which also reduces contention on shared resources by recording removed nodes locally for each thread.
When compared with other memory reclamation algorithms (e.g., Hazard Pointers~\cite{hazardpointers}, StackTrack~\cite{stacktrack}), DEBRA demonstrates lower overhead and is applicable to many
data structures~\cite{debra}.

\textbf{EBR Overview.}
EBR guarantees that unreachable objects are freed by maintaining a collection of references to recently retired objects. It operates under the assumption that threads cannot save references to objects outside of the scope of an operation (i.e., during quiescence).To ensure that an object can be freed without problems, EBR monitors the epoch observed by each thread and the objects retired during each epoch. The epoch is only incremented after all active threads have announced that they have observed the current epoch value. When a new epoch is started, any objects retired two epochs prior can be safely freed.

%

%
%

%
\textbf{Freeing Data Structure Nodes.} 
EBR guarantees that no node is freed while concurrent range queries (as well as any concurrent primitive operation) may access it; 
and, bundling guarantees that no range query that starts after physically removing a node will traverse to this node.
As an example, consider the two following operations: \textit{i}) a range query, $R$, whose range includes node $x$; and \textit{ii}) a removal operation, $U_t$, which is linearized at time $t$ and removes $x$.
If $R$ is concurrent with $U_t$, then EBR will guarantee that $x$ is not freed since $R$ was not in a quiescent state and a grace period has not passed.
In this case, $R$ may safely traverse to $x$ based on its observed timestamp, without concern that the node may be freed.
On the other hand, if $R$ starts after $U_t$, then trivially $x$ will never be referenced by $R$ and is safe to be reclaimed since $R$ observed a timestamp greater than or equal to $t$.

\textbf{Freeing Bundle Entries.} Bundle entries are reclaimed in two cases. The first, trivial, case is that bundle entries are reclaimed when a node is reclaimed. The second case is more subtle. After a node is freed, there may still exist references to it (in other nodes' bundles) that are no longer necessary and should be freed.
Bundle entries that have a timestamp older than the oldest active range query can be retired only if there also exists a more recent bundle entry that satisfies the oldest range query.
This cleanup process may be performed during operations themselves or, as we implement, delegated to a background thread.

To keep track of active range queries we augment the global metadata with \texttt{activeRqTsArray}, an array of timestamps that maintains their respective starting timestamp. 
During cleanup, this array is scanned and the oldest timestamp is used to remove outdated bundle entries. 
Reading the global timestamp and setting the corresponding slot in \texttt{activeRqTsArray} must happen atomically to ensure that a snapshot of the array does not miss a range query that has read the global timestamp but not yet announced its value.
This is achieved by first setting the slot to a pending state, similar to they way we protect bundle entries, which blocks the cleanup procedure until the range query announces its starting timestamp.
Second, the cleanup thread has to be protected by EBR as well, just like other operations. 

\begin{table}[h]
    \small
    \centering
    \begin{tabular}{c|c|c|c|c|c|c}
        & & \multicolumn{5}{c}{\textbf{Update \%}} \\
        \hline
        && \textbf{0\%} & \textbf{10\%} & \textbf{50\%} & \textbf{90\%} & \textbf{100\%}  \\
        \hline
        \parbox[t]{2mm}{\multirow{4}{*}{\rotatebox[origin=c]{90}{\textbf{Delay ($d$)}}}} & $0ms$ & 4 & 11 & 14 & 12 & 13 \\
        & $1ms$ & 0 & 10 & 13 & 10 & 13 \\
        & $10ms$ & 2 & 9 & 11 & 7 & 8 \\
        & $100ms$ & 0 & 8 & 8 & 0 & 0 \\
    \end{tabular}
    \caption{\% overhead when enabling memory reclamation. }
    \label{tab:reclamation}
\end{table}

\textbf{Experiments.}
Instead of repeating experiments, we focus on performance relative to the leaky bundled data structures while adjusting the delay ($d$) between cleanup iterations, as shown in Table~\ref{tab:reclamation}.
For mixed and update heavy workloads, freeing nodes and bundle entries entails a maximum of 14\% degradation in performance with an aggressive cleanup delay ($d=0$).

\end{document}